\documentclass[floatfix,prd,lengthcheck,showpacs,amssymb,amsmath,amsfonts,aps,altaffilletter,nofootinbib,nopreprintnumbers,showpacsm]{revtex4-1}

\usepackage[utf8]{inputenc}
\usepackage[dvipsnames]{xcolor}
\usepackage{graphicx}
\usepackage{subfig}
\usepackage{amsmath}
\usepackage{acronym}
\usepackage{multirow}
\usepackage{tabu}
\usepackage{import}
\usepackage{caption}
\usepackage{hyperref}
\usepackage{dsfont}
\usepackage{booktabs}
\usepackage{siunitx}
\usepackage{lineno}   % For line numbering
\usepackage{ulem}
\usepackage{diagbox}

\hypersetup{
    colorlinks=true,
    linkcolor=blue,
    filecolor=magenta,      
    urlcolor=cyan,
}

\begin{document}
\newcommand{\mlp}[1]{\textcolor{black}{#1}}
\title[]{Ameliorating transient noise bursts in gravitational-wave searches for intermediate-mass black holes}
\def\thefootnote{\textdagger}\footnotetext{Corresponding author: \href{mailto:melissa.lopez@ligo.org}{melissa.lopez@ligo.org}}\def\thefootnote{\arabic{footnote}}

\author{Melissa Lopez$^{1,2, \text{\textdagger}}$}
\author{Giada Caneva Santoro$^3$}
\author{Ana Martins$^4$}
\author{Stefano Schmidt$^{1,2}$}
\author{Jonno Schoppink$^2$}
\author{Wouter van Straalen$^2$}
\author{Collin Capano$^{5, 6}$}
\author{Sarah Caudill$^{6, 7}$}

\affiliation{$^1$Nikhef Science Park 105,
1098 XG, Amsterdam, The Netherlands.}
\affiliation{$^2$Institute for Gravitational and Subatomic Physics (GRASP) Utrecht University, Princetonplein 1, 3584 CC, Utrecht, The Netherlands.}
\affiliation{$^3$Institut de Fisica d’Altes Energies (IFAE), Barcelona Institute of Science and Technology, E-08193 Barcelona, Spain}
\affiliation{$^4$Institute of Theoretical Astrophysics, University of Oslo, Sem Sælands vei 13, 0371 Oslo, Norway.}
\affiliation{$^5$Department of Physics, Syracuse University, Syracuse, NY 13244, USA}
\affiliation{$^6$Department of Physics, University of Massachusetts, Dartmouth, MA 02747, USA}
\affiliation{$^7$Center for Scientific Computing and Data Science Research, University of Massachusetts, Dartmouth, MA 02747, USA}

% Please add ORCIDs here in case we submit to a journal which accepts them
% Gareth: 0000-0002-0355-5998
% Alex: 0000-0002-1850-4587
% Marton: 0000-0002-5354-5683

\date{\today}

\begin{abstract}
 The direct observation of intermediate-mass black holes (IMBH) populations would not only strengthen the possible evolutionary link between stellar and supermassive black holes, but unveil the details of the pair-instability mechanism and elucidate their influence in galaxy formation. Conclusive observation of IMBHs remained elusive until the detection of gravitational-wave (GW) signal GW190521, which lies with high confidence in the mass gap predicted by the pair-instability mechanism. 
Despite falling in the sensitivity band of current GW detectors, IMBH searches are challenging due to their similarity to transient bursts of detector noise, known as glitches. In this proof-of-concept work, we combine a matched-filter algorithm with a Machine Learning (ML) method to differentiate IMBH signals from non-transient burst noise, known as glitches.
In particular, we build a multi-layer perceptron network to perform a multi-class classification of the output triggers of matched-filter. In this way we are able to distinguish simulated GW IMBH signals from different classes of glitches that occurred during the third observing run (O3) {in single detector data}.
{We train, validate, and test our model on O3a data, reaching a true positive rate of over $90\%$ for simulated IMBH signals. \mlp{To test the generalization ability over the evolutionary observing run, we test on the useen data of O3b, which yields a true positive rate of over $70\%$} . We also combine data from multiple detectors to search for simulated IMBH signals in real detector noise, providing a significance measure for the output of our ML method.}
%Interferometer data contains numerous noise transients, or “glitches”, that can mimic true gravitational waves, reducing the sensitivity of the match-filtering search by increasing the rate at which random coincidences occur. High-mass binary black hole mergers are particularly susceptible since they resemble short noise transients. We show that we can use a template bank of compact binary gravitational waveforms as a probe of the frequency evolution of transients in the data. We use the GstLAL detection pipeline to produce matched-filter triggers, under the assumption that short high-mass black hole signals and glitches yield different patterns of GstLAL triggers in time and template parameters. We propose an inexpensive statistic, derived from the Random Forest algorithm, based on the triggering patterns, that can easily distinguish between real events and noise, consequently increasing the significance of gravitational-wave candidates.
\end{abstract}

\maketitle

\section{Introduction}
\label{sec:intro}

\subsection{Motivation}

{As stellar evolution models predict, stars with helium core masses in the range $\sim 32-64\,\text{M}_{\odot}$ are subject to pulsation pair instability, while stars with helium core masses in the range $\gtrsim 50 - 130\,\text{M}_{\odot}$ leave no remnant due to pair-instability supernovae (PISN) \cite{Woosley:2007qp, Woosley:2021xba}. Nonetheless, we find intermediate-mass black holes (IMBH) within this mass gap and beyond, spawning from  $\sim 10^2 - 10^5\,\text{M}_{\odot}$.}

{Although Advanced LIGO \cite{LIGOScientific:2014pky} and Advanced Virgo \cite{VIRGO:2014yos} gravitational-wave (GW) detectors detected 11 candidates during the first observing run (O1) and the second observing run (O2)
\cite{LIGOScientific:2018mvr, LIGOScientific:2019ysc, LIGOScientific:2021usb}, 
the detection of IMBHs in GW searches remained elusive until the detection of GW190521 during the third observing run (O3) \cite{LIGOScientific:2020ibl, LIGOScientific:2021tfm}. The estimate of the individual {source-frame} mass components of GW190521 was $(m_1, m_2) = (85^{+11}_{-14}, 66^{+17}_{18})\, \text{M}_{\odot}$, with a final remnant mass of $M_{f} = 142^{+28}_{-16}\, \text{M}_{\odot}$ making it the first conclusive direct observation of an  IMBH \cite{PhysRevLett.125.101102}. Final analyses of O3 data yielded even higher mass IMBH GW events. In particular, 
{GW190426\_190642~\cite{LIGOScientific:2021usb} supersedes GW190521 with a final mass of $M_{f} = 172.9^{+37.7}_{-33.6}\, \text{M}_{\odot}$, while GW200220\_061928~\cite{LIGOScientific:2021djp} has probably the highest final mass of the second half of O3 (O3b) $M_{f} = 141^{+51}_{-31}\, \text{M}_{\odot}$.}}

{The origin of supermassive black holes (SMBHs) of masses beyond $10^5\,\text{M}_{\odot}$ remains a fundamental mystery, despite their presence in nearly every galaxy, including the Milky Way \cite{Richstone:1998ky, EHCSgtA*}. Recent observational surveys with the James Webb Telescope~\cite{Gardner:2006ky} %have identified several hundreds of luminous quasars at redshifts $z > 6-7$, indicating the existence of SMBHs with masses on the order of $10^9 - 10^{10}\,\text{M}_{\odot}$ at such early cosmic epochs \cite{Suh:2024jbx}. These findings
challenge models of SMBH formation and growth, raising questions about the origin of seed black holes and the mechanisms enabling their rapid growth to supermassive scales. 
The PISN mass gap implies that stellar collapse alone may not account for SMBH formation. A plausible scenario is the hierarchical mergers of IMBHs, providing a potential evolutionary bridge between stellar-mass black holes and SMBHs. \cite{2020ARA&A..58..257G, LIGOScientific:2021tfm}. 
Directly observing IMBH populations would not only strengthen the proposed evolutionary link and unveil the role of the PISN mechanism, but also serve as a rigorous test of General Relativity, given the characteristically strong merger and ringdown signals they produce \cite{Isi2019HierarchicalTO}.}

Despite falling in the sensitivity band of current GW interferometers, GW signals from IMBH mergers are challenging as few cycles of the signal can be observed with current ground-based detectors. {Initial, GW IMBH searches were restricted to using the model waveform-independent algorithm coherent Wave Burst (cWB)\cite{PhysRevD.72.122002, SKlimenko_2006, PhysRevD.83.102001} and a ringdown templated search \cite{Aasi_2015}  in order to probe the merger-ringdown phase.} {The improvement of detector sensitivity  at low frequencies in the advanced era allowed to probe the short inspiral phase with matched-filtering techniques.} Nowadays, state-of-the-art searches employ {the modeled-independent cWB in its IMBH configuration} \cite{Klimenko:2015ypf, Gayathri:2019omo, Drago:2020kic}, and  matched-filter-based techniques using modeled-waveforms \cite{Allen:2005fk, Messick:2016aqy, usman2016pycbc, Sachdev:2019vvd,  Chandra:2021wbw, Chandra:2021xvs, Tsukada:2023edh}. 

\begin{figure}[]
\centering
\includegraphics[width=0.5\textwidth]{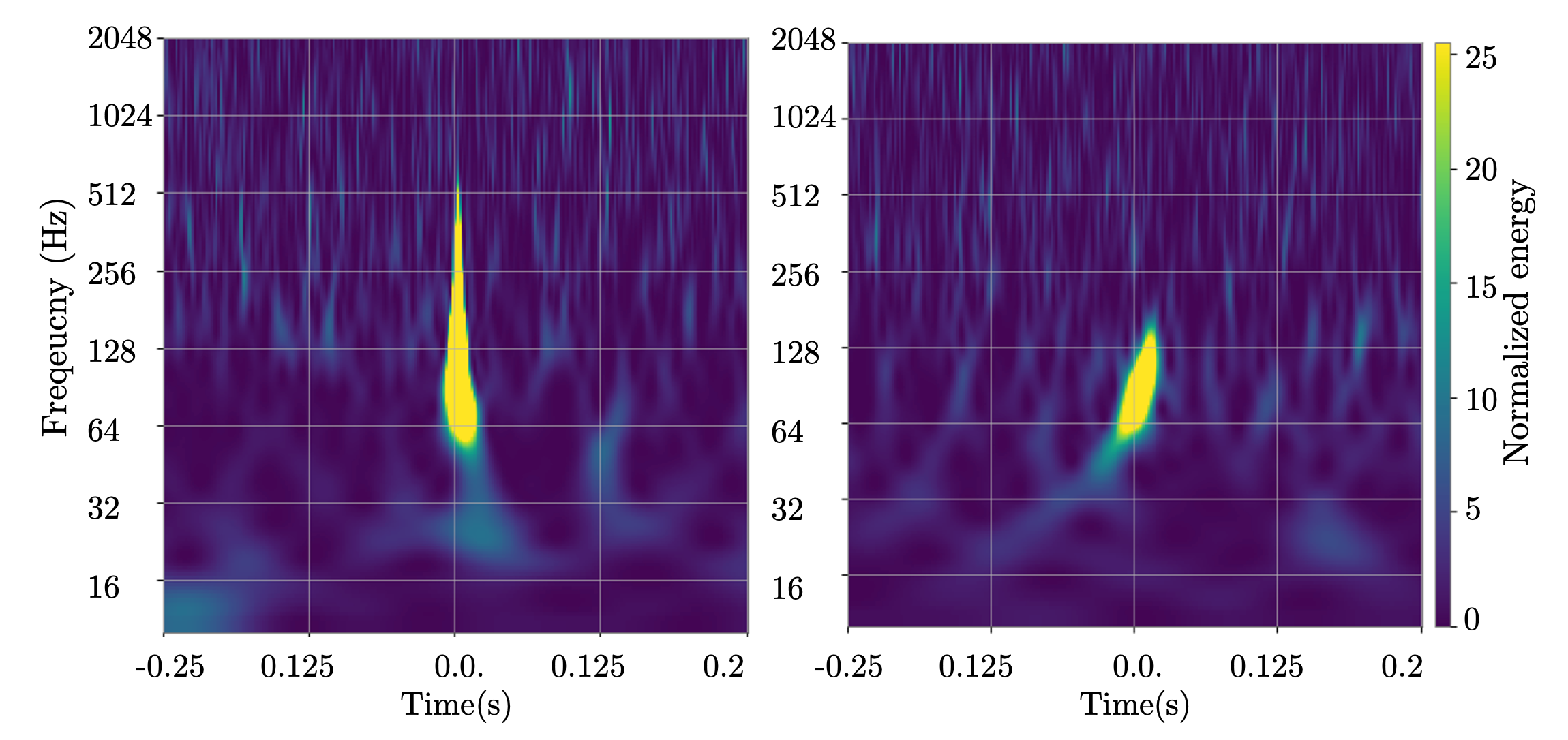}
\caption{\textit{(Left)} Q-transform of a \texttt{Blip} glitch retrieved from \textit{Gravity Spy} \citep{zevin2017gravity}. \textit{(Right)} Q-transform of a gravitational wave (GW) signal from the merger of two high mass black holes to form an IMBH with total mass $106.6^{+13.5}_{
-14.8} M_{\odot}$.}
\label{fig:glitch_IMBH}
\end{figure}

Despite the improvement in sensitivity, IMBH searches are still hampered by non-Gaussian transient burst noise, known as glitches, that can mask or mimic GW candidates, reduce the amount of analyzable data increasing the noise floor and affect the estimation of the power spectral density (PSD) \cite{LIGOScientific:2017tza, Virgo:2022ysc, Glanzer:2022avx}. {As an example, in} Fig. \ref{fig:glitch_IMBH} we show the similarities between a \texttt{Blip} glitch and an IMBH signal. {\texttt{Blips}
 are short glitches ($0.2\,$s) of unknown origin that have a characteristic morphology of a symmetric ‘teardrop’ shape in time-frequency in the range [30,250] Hz. As we can observe in Fig. \ref{fig:glitch_IMBH}, \texttt{Blips} have a similar duration to IMBH signals and intersect the frequency band of these signals.} 
 
 To enhance the sensitivity of current searches, a interesting approach is to combine them with Machine Learning (ML) algorithms, fostering a synergistic relationship.

\subsection{Previous work}

 {Through the years, scientists have implemented multivariate ML methods to enhance conventional searches, allowing them to more accurately distinguish GW signals from background glitches. The implemented ML methods would learn a single feature vector derived from multi-detector analysis to perform the binary classification task. Such applications have proven successful in the context of binary black holes \cite{Baker:2014eba, Hodge:2014yca}, gamma-ray bursts \cite{Adams:2013pna, Kim:2014nba}, and burst searches \cite{Mishra:2021tmu, Szczepanczyk:2022urr, Gayathri:2020bly, Lopez:2021ikt}. Nonetheless, these multivariate methods are not restricted to a single feature vector; they can also integrate information from several single-detector analysis~\cite{Kapadia:2017fhb} and other data representations, such as singular value decomposition  
 ~\cite{2024arXiv241015513M} to distinguish GW signals from background glitches.}

\subsection{Relation to previous work}

{
It may be possible to construct a more comprehensive data structure that fully captures the unique features separating GW signals from glitches. While earlier approaches rely on a single feature vector \cite{Baker:2014eba, Hodge:2014yca, Adams:2013pna, Kim:2014nba, Mishra:2021tmu, Szczepanczyk:2022urr, Gayathri:2020bly, Lopez:2021ikt}, our method instead utilizes sets of feature vectors derived from matched-filtering. This process correlates the input detector data with collections of modeled GW signals—referred to as \textit{templates}—which are organized into \textit{template banks}.}

{The key idea is that when we ``match" the unknown detector data $s(t)$ with the templates of a bank, a handful of templates, say $N$, will be sufficiently similar to $s(t)$ within a finite time window $\Delta t$. Associated to each matching template, the search pipeline will launch several ``triggers", all flagging the same potential candidate.}
We refer to this inherent cluster structure  as \textit{cluster track}, or simply \textit{tracks}, {inspired by the tracking produced in particle colliders \cite{Shiltsev2020ModernAF}}.
{Given the morphological time-frequency differences between glitches and GW signals, we expect these ``{clusters of triggers}" to exhibit a meaningful structure to differentiate them with ML methods. }
{We do so by introducing a new statistic: a multivariate multi-class probability vector, which measures the probability that each track is generated by either an astrophysical signal or by one of the glitch classes considered. }

{In this work we use the tracks generated in single-detector, right after the matched-filtering step. Hence, to maintain as much information as possible, we reproduce the IMBH search of GstLAL during O3 using LIGO Hanford (H1), LIGO Livingston (L1) and Virgo (V1) \cite{LIGOScientific:2021tfm}. With this information, the ML algorithm learns to distinguish glitches from simulated IMBH signals in a controlled environment.}
Afterwards, we use the algorithm to detect a set of simulated IMBH, performing a background estimation. This proof-of-concept study demonstrates that even with a restricted template bank, it is possible to effectively separate signals from glitches, making a potential impact on the accuracy of GW analysis.  It is relevant to to note that while this particular investigation focuses on IMBH, this method can be extended to other compact binary coalescence  (CBC) signals.

\textit{Outline of the paper:} In section \ref{sec:gstlal} we introduce the current state-of-the-art matched filtering technique and its generated tracks.  In section \ref{sec:dataset} we provide an overview of the different types of signals employed and the details regarding the construction of the data sets. In section \ref{sec:methodology} we describe the multivariate ML model, its input and the learning procedures employed to enhance its performance. In section \ref{sec:results} we show the training and  evaluation metrics of the model, as well as the search of simulated IMBH signals. 
Finally, in section \ref{sec:conclusions} we conclude and propose avenues for future research.

\section{Matched Filtering pipeline}\label{sec:gstlal}

As the typical amplitude of a GW signal $h(t)$ is orders of magnitude smaller than the detector background noise $n(t)$, sophisticated detection algorithms, or pipelines, are needed to identify CBC signals. 
Most of them are based on matched-filter, which in turn relies on the precise models of the sources, for instance, employing Post-Newtonian approaches \cite{PhysRevLett.74.3515} or the effective-one-body formalism \cite{Buonanno:1998gg}. Some examples are the GstLAL-based inspiral~\cite{2017Messick, Sachdev:2019vvd, hanna2020fast, cannon2021gstlal, Tsukada:2023edh}, PyCBC~\cite{Allen:2005fk,allen2005chi, dal2014implementing, usman2016pycbc, nitz2017detecting, davies2020extending, nitz2018rapid}, MBTA~\cite{adams2016low, aubin2021mbta} and SPIIR~\cite{chu2022spiir} pipelines. In this  work, we use the GstLAL-based inspiral pipeline, hereafter refered to as GstLAL, so in the following sections we provide a high-level overview of some of its particularities. 

\subsection{Time-domain matched filtering}\label{sec:matchedfilter}

{The tasks of a matched-filter search} is to find a target signal $h(t)$ given the detector output $s(t) = n(t) + h(t)$. This is achieved by  cross-correlating $s(t)$ with a modelled waveform $h_{m}(t)$, known as a template. We can define a time-dependent complex scalar product as \cite{Allen:2005fk},
\begin{equation}
\label{eq:wiener}
\langle s|h \rangle(t) =  2\int_{-\infty}^{\infty} \frac{\tilde{h}_m^{*} (f) \tilde{s}(f)}{S_{n}(f)}e^{i 2 \pi f t} df, 
\end{equation}
where $S_{n}(f)$ represents the one-sided PSD, and $\tilde{\cdot}$ and $^*$ denote the Fourier transform and the complex conjugate, respectively.
We can also define the real and imaginary part of the scalar product as ${\langle s|h \rangle(t) = (s|h)(t) + i[s|h](t)}$, where we define ${(s|h)(t) = \Re \langle s|h \rangle(t)}$ and ${[s|h](t) = \Im \langle s|h \rangle(t)}$.
With this definition and given $h(t)$,  its normalized time series is $\hat{h} = h /(h|h)$.

The GstLAL analysis performs matched filtering directly in the time domain \cite{Sachdev:2019vvd, 2017Messick}. In accordance with the notation of \cite{2017Messick}, and  given a time series $s(t)$ and a normalized complex template $h_{c}(t)$, the output of the matched-filter technique is a complex time series,
\begin{equation}
\label{eq:complex_snr}
z(t) = (s|h_{\mathcal{R}})(t) + i(s|h_{\mathcal{I}})(t),
\end{equation}
where $h_{\mathcal{R}}$ and $h_{\mathcal{I}}$ are the real and imaginary part of $h_{c}(t)$.
The absolute value $|z(t)|$ of the complex matched filtering as the detection statistics is usually referred to as signal-to-noise ratio (SNR),
\begin{equation}
\label{eq:snr}
\rho(t) = |z(t)| = \sqrt{(s|h_{\mathcal{R}})^2(t) + i(s|h_{\mathcal{I}})^2(t)}.
\end{equation}
The SNR quantifies the ``agreement" of the data with a template and as such, it constitutes the main detection statistics. A detection can be claimed only if the SNR exceeds a certain threshold.

\subsection{Template banks}

CBC template waveforms depend on the intrinsic parameters  of the source $\lambda_{int}$, such as the masses ($m_{1}, m_{2}$) and z-spins ($s_{1z}, s_{2z}$) of the binary components, and on extrinsic parameters $\lambda_{ext}$, which are related to the position of the source concerning the observer, such as the luminosity distance $D$. Thus, the goal of matched filtering is to maximize the detection statistic, SNR, over all these parameters.
Luckily, for aligned-spin signals, the effect of the extrinsic parameters can be absorbed into a constant scale factor and a phase~\cite{Creighton:2011zz} and the SNR (Eq.~\ref{eq:snr}) maximizes over such extrinsic parameters.
{Therefore, we are only left with the} maximisation over the intrinsic parameters {which} is usually performed by a brute force approach, where the SNR time-series is computed for a dense grid of {modelled GW} signals, called {\it templates}{, characterized by $\lambda_{int}$}. %Each template is characterized by the masses $m_1, m_2$ and the z-spins $s_{1z}, s_{2z}$ of the signal it represents. 
Templates are gathered in large template banks, which are usually generated \cite{Ajith:2012mn} to balance between a small loss of SNR due to the discreteness of the grid and a manageable computational cost.

 Different pipelines employ different strategies to cross-correlate template banks with the output data from the detector while mitigating glitches. 
{In this work we are interested {in} utilizing the product of this cross-correlation targeting IMBH signals, which are heavily harmed by glitches due to their short duration and similar frequency range. For this aim, we employ GstLAL to reproduce the IMBH search of O3 \cite{LIGOScientific:2021tfm}. We use the IMBH template bank which  uses $44902$ templates to cover aligned-spin systems with total mass $M = m_1 +m_2 $ in range $[50, 600] \, \text{M}_{\odot}$ and mass ratio $q=\frac{m_1}{m_2}$ between $[1, 10]$, with both spins in the range $[-0.98, 0.98]$}. The starting frequency of the analysis is at $10\,$Hz. For further details the interested reader can refer to \cite{LIGOScientific:2021tfm}. 

\subsection{$\xi^2$-statistic}

The detector strain contains glitches that can mask or mimic GW signals, producing large peaks in the SNR time series. As {glitches in the detector could yield large SNRs, to mitigate them} GstLAL also calculates a signal consistency check, known as $\xi^2$-statistic, whenever it records an SNR above a certain threshold. This check is performed by determining the similarity between the SNR time series of the data and the expected SNR time series from the real signal within a \( \delta t \) time window around the peak \cite{2017Messick}. Mathematically, the $\xi^2$-statistic is defined as,

\begin{equation}
\label{eq:chisq}
\xi^2 = \frac{\int_{-\delta t}^{\delta t}|z(t) - z(0)R(t)|^2 dt}{\int_{-\delta t}^{\delta t} (2-2|R(t)|^2)dt }, 
\end{equation}
where $z(t)$ is the complex SNR time series, $z(0)$ is its peak and $R(t)$ is the auto-correlation series between the complex template waveform and itself.

\subsection{Triggers}\label{sec:triggers}

{When the SNR associated with a template  exceeds a certain threshold, the GstLAL pipeline records a feature vector known as a ``{trigger,}" which contains the maximum SNR, $\xi^2$, masses, spins, and other parameters.}

It often happens that multiple triggers from different templates match the same signal in the data (either of terrestrial or astrophysical origin). They differ in the trigger time, as well as SNR and $\xi^2$ values.
All such triggers must be associated with the same candidate and for this reason, they need to be clustered together to form a single candidate.

\begin{figure}[h]
    \centering
    \includegraphics[width=0.5\textwidth]{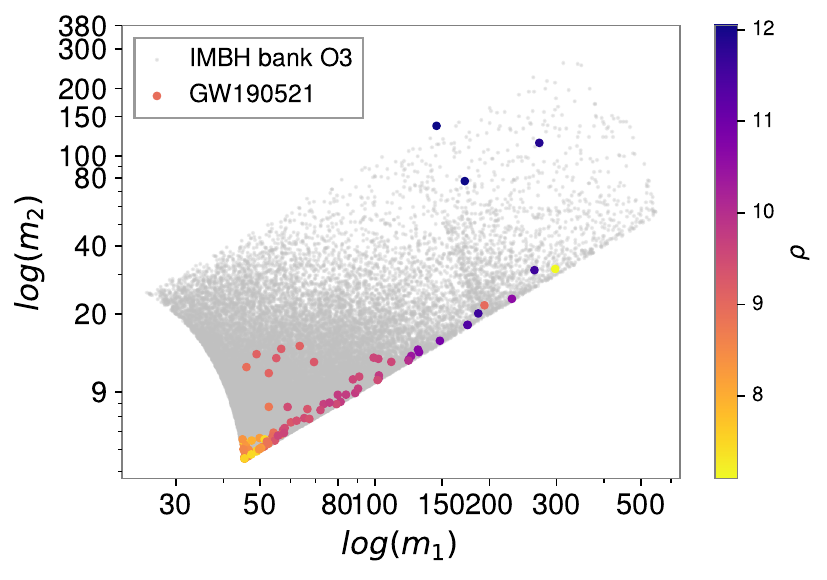}
    \caption{IMBH template bank from O3 (grey) as a function of the progenitors' masses measured in $M_{\odot}$ and logarithmic scale. In colour we show  the maximum SNR $\rho$ of the ``matching" templates of GW190521 in LIGO Livingston (L1).}
    \label{fig:gw190521snr}
\end{figure}

The GstLAL pipeline adopted a simple approach to solve this problem: {the triggers with the largest SNR peak within the vicinity ($\pm \SI{1}{s}$) will be defined as the cluster centroid and utilized for further analysis}\footnote{Another clustering of triggers is performed at a later stage of the pipeline. However, this has no importance for our purposes.}. While this approach has the benefits of being simple, it discards many pieces of information, such as the position of the various triggers within the template bank and their time ordering.
{For illustration, Fig. \ref{fig:gw190521snr} shows the tracks that ``matched" GW190521 during the IMBH search in O3 for LIGO Livingston (L1).} We represent the IMBH template bank (grey) as a function of the progenitor masses, and the track is coloured as a function of the maximum SNR $\rho$.

Afterwards, each trigger is ranked \cite{Cannon:2015gha, Hanna:2019ezx, Tsukada:2023edh} according to its probability {of originating from an astrophysical signal}. The goal of the pipeline is then to obtain a list of triggers, ordered by their likelihood $\Lambda$ to be of astrophysical origin.

\section{Data}\label{sec:dataset}

%In Section \ref{sec:matchedfilter} we provided an overview of the basic matched-filtering method, and in \ref{sec:triggers} we described how a trigger is generated by the matched-filtering algorithm used in this research, GstLAL, when a template waveform is cross-correlated with a signal. 

%It often happens that different templates ``{match}" an {unknown signal buried within the} time series $s(t)$. {As a result,} $N$ triggers are {``{matched}" or associated} with {such unknown signal within a finite time window $\Delta t$, which we refer to as \textit{cluster} of triggers}. {After recording all such triggers, GstLAL selects the loudest SNR trigger within the cluster (clustering step), which we refer to as the \textit{centroid}. Afterwards, these centroids are ranked according to GstLAL likelihood function $\Lambda$ to claim a candidate \cite{Cannon:2015gha}. }

%{Given the morphological time-frequency differences between glitches and GW signals, we expect the ``{matched}" templates to exhibit a meaningful structure to differentiate them with ML methods. We refer to this inherent structure as \textit{cluster tracks}, or simply \textit{tracks}. In particular, we are interested in {tracks generated in} single-detector, right after the matched-filtering step. Hence, t}o maintain as much information as possible, we reproduce the IMBH search of GstLAL during O3~\cite{LIGOScientific:2021tfm}, terminating the process before the clustering step. 

Once we have collected the tracks data  we aim to utilize a supervised ML method to differentiate GW from glitches. Thus, we construct two different data sets: a \textit{controlled} data set which contains well-known glitches and  simulated IMBH signals, referred to as \textit{known data set}; and a second data set which contains real GW signals and other \textit{a priori} unknown signals, {which we use to construct an accidental background of time shifts (see Section \ref{sec:timeshift} for details)}, referred to as \textit{unknown data set}. While the first data set is employed to assess the performance of the method, the second data set is used to understand the significance of the ML statistic. 

In Section \ref{sec:glitches}, we describe the simulated IMBH signals, and {the different glitch classes} {present in the \textit{known data set}}.  In Section \ref{sec:clustering} we describe {the} clustering procedure {utilized for defining a track}. In Section \ref{sec:features} we show the patterns generated by an IMBH simulation and a glitch through the template bank, and we define the feature vector associated with the triggers that will be the input to the multivariate ML model. {Lastly, in Section \ref{sec:timeshift} we discuss how we construct the accidental background.}

\subsection{Injections and glitches}
\label{sec:glitches}

Here, we describe the simulated IMBH waveforms, and the six different glitch populations 
from the data set of \textit{Gravity Spy} during O3 in H1, L1 and V1 \cite{zevin2017gravity, Glanzer:2022avx}.  As \textit{Gravity Spy} is a classification algorithm, it returns a classification for each glitch. Glitches have been selected with a \textit{Gravity Spy} confidence level greater than $90\%$, indicating that the model is highly confident that the glitches belong to the assigned class.

\begin{figure*}[ht]
    \centering
    \includegraphics[width=\textwidth]{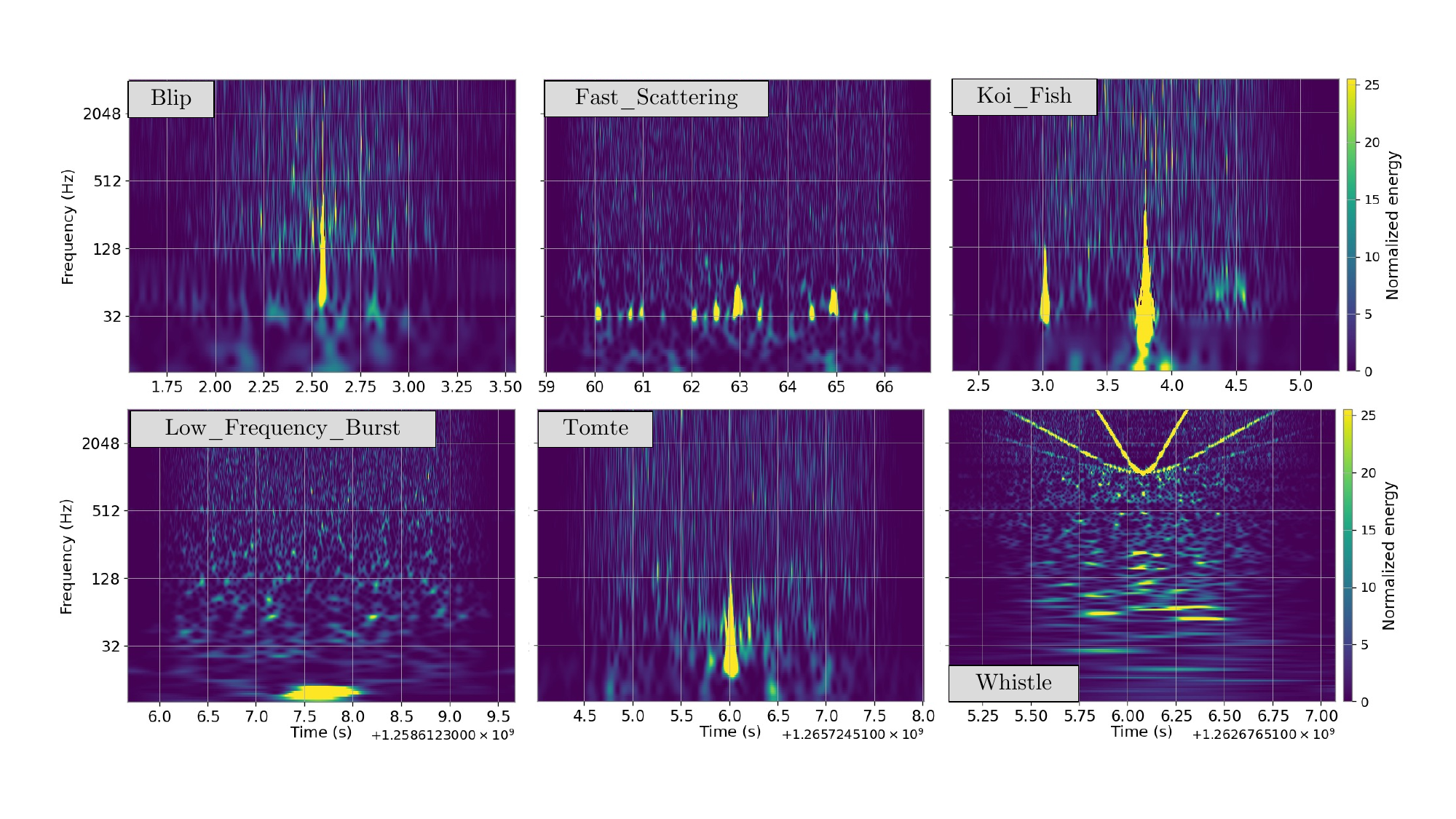}
    \caption{Q-transform of GW190521 and {example} of different glitch classes selected from \textit{Gravity Spy} {at $> 90 \%$ confidence }in LIGO Livingston  (L1)~\cite{zevin2017gravity}.
    }
    \label{fig:glitches}
\end{figure*}

The chosen glitch classes encompass both short-duration (microsecond) and long-duration (second) signals, which are commonly observed in current ground-based interferometers and cover a broad frequency range. We provide a detailed description of each class below and illustrate the various glitch morphologies in Fig. \ref{fig:glitches}.

\begin{itemize}

\item \texttt{Injections}:
For this aim we use \texttt{IMRPhenomD} approximant~\cite{Khan:2015jqa} with masses $m_{1} \in [50, 400]\,\text{M}_\odot$ and $m_{2} \in [10, 250]\,\text{M}_\odot$, and dimensionless spins $\chi_1, \chi_2 \in [0, 0.99]$, uniformly sampled within the specified range {with aligned spins}. The distances are uniformly sampled from the range $d \in [10, 250]\,$Mpc. The inclination is sampled from a uniform distribution, and the sky-localization is randomized.

\item \texttt{Blip}: these glitches have a characteristic morphology of a symmetric ``teardrop" shape in time-frequency in the range $[30, 250]\,$Hz with short-durations, $\sim 0.04\,$s \cite{Cabero:2019orq}. 
Due to their abundance and form, these glitches hinder both the unmodeled burst and modelled CBC can searches with particular emphasis on compact binaries with large total mass { due to the short duration of their waveforms,} 
among others  \cite{Abbott_2018, abbott2016characterization}. Moreover, as their source is still unknown, they cannot be removed from astrophysical searches.

\item \texttt{Fast$\_$Scattering}:  they appear as short-duration arches ($\sim 0.2 - 0.3\,$s) in the frequency range $[20 -60] \,$Hz. These glitches are strongly correlated with ground motion in range $[0.1 -0.3] \,$Hz and $[1-6] \,$Hz, which in turn is associated with thunderstorms and human activity near the detector. Note that this class was added during O3, and it is more abundant in L1 than H1, but not present in Virgo, due to differences in ground motion and detector sensitivity \cite{Soni:2021cjy}. 

\item \texttt{Koi$\_$Fish}: their naming comes from their imaginative frontal resemblance to koi fishes. They are also similar to \texttt{Blip} but typically feature higher-SNR than the latter,
 spanning the frequency range of $\sim 20 - 1000\,$Hz.

\item \texttt{Low$\_$Frequency$\_$Burst}: they are short-duration  glitches ($\sim 0.25\,$s) in the frequency range $[10 - 20]\,$ Hz with a distinctive {hump shape at the bottom of spectrgrams}.
 These occurrences were prevalent in L1 data during O1 and H1 data in O3a.

\item \texttt{Tomte}: these glitches are also short-duration ($\sim 0.25\,$s) with a characteristic triangular morphology. As \texttt{Blip} glitches, their source is unknown, so they cannot be removed from astrophysical searches. 

\item \texttt{Whistle}: these glitches have a characteristic V, U or W shape at higher frequencies ($\gtrsim$ 128 Hz) with typical durations $\sim 0.25\,$s.  {They are caused when radio-frequency signals beat with the voltage-controlled oscillators \cite{Nuttall:2015dqa}.} 

\end{itemize}

\subsection{Clustering}
\label{sec:clustering}

 In this study, we are interested in exploring the generalization power between the first and second half of the third observing run, namely O3a and O3b, respectively. For this aim, we use the \textit{known data set} from O3a for training, validation and testing, while the \textit{known data set} of O3b is used to test this generalization power. Afterwards, we use the unknown data set to construct an accidental background (see \ref{sec:timeshift}) and assess the significance of our ML statistic.
 
In the following, we describe how the cluster tracks of these data sets were defined.

\begin{figure}[!h]
    \centering
    \includegraphics[width=0.5\textwidth]{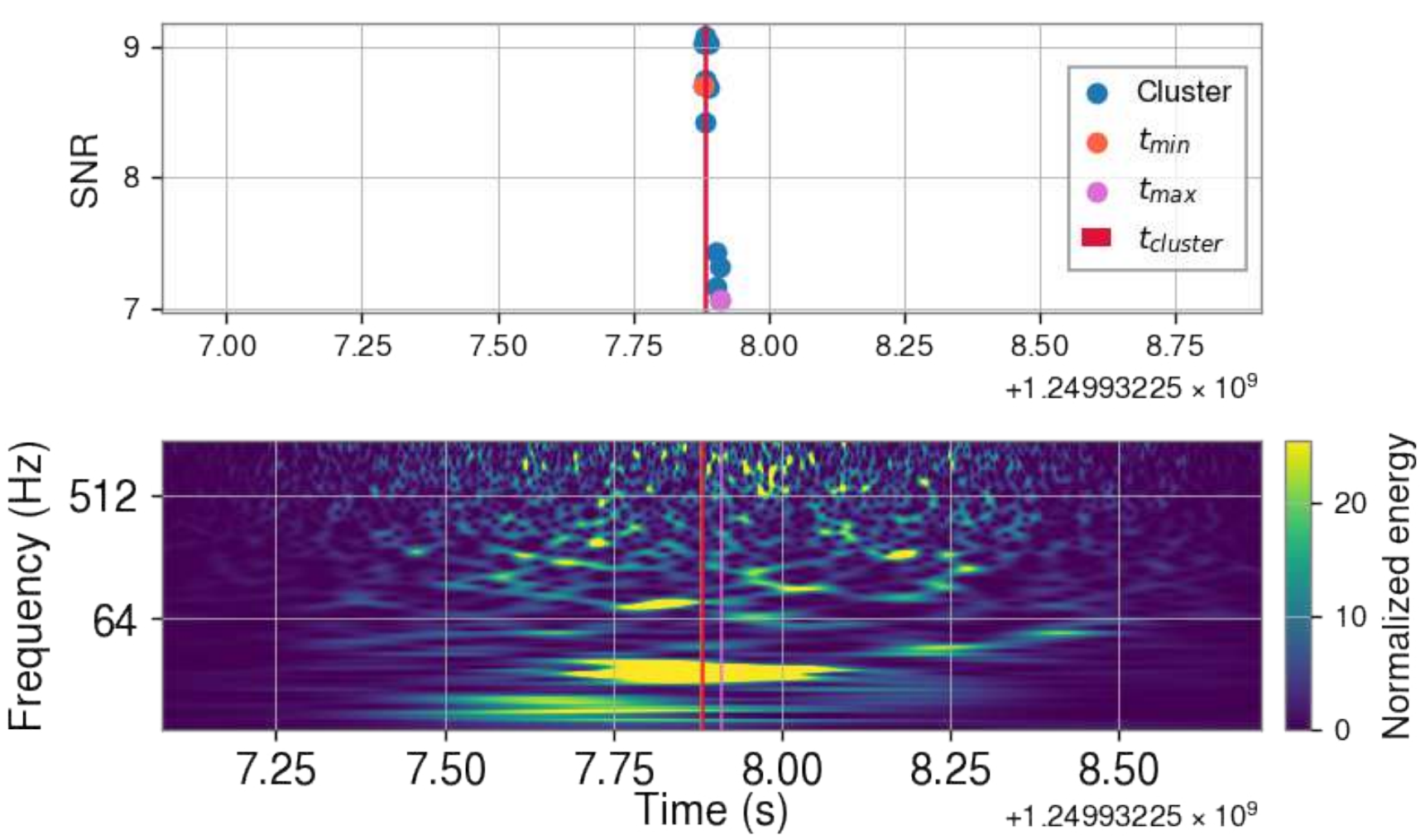}
    \caption{ A \texttt{Fast Scattering} glitch labelled by \textit{Gravity Spy}. \textit{(Top)} SNR of GstLAL tracks as a function of time, where each point represents a template. $t_{min}$ (orange) and $t_{max}$ (pink) mark the beginning and end of the cluster, respectively, while $t_{cluster}$ (red) indicates its centroid given by \textit{Gravity Spy}. \textit{(Bottom)} Q-transform of the time series containing the \texttt{Whistle} labelled by \textit{Gravity Spy}.
    }
    \label{fig:offset_simple}
\end{figure}

\begin{itemize}
\item[-] \textbf{Known data set}: Since we know the GPS time when an \texttt{Injection} or a glitch occurred, we use it as {the} centroid {of the cluster},  $t_{cluster}$, and we select a window $t_{w} = \pm\, \Delta t$ around it. We experimented with $\Delta t \in \{0.05, 0.1, 0.2, 0.5, 1\}\,$s to find the optimal $t_{w}$ of the cluster{--these were chosen to be of length comparable to typical duration of an IMBH signal.} It is  relevant to note that while we experiment with different $t_{w}$, once the algorithm has trained with a given $t_{w}$, this value is fixed for validation and testing.  {We provide a thorough discussion on the selection of optimal $t_{w}$ in Section \ref{sec:tw}.} %It is relevant to note that while $t_{cluster}$ is the ground truth for injected IMBH signals, it might not be the case for glitches.  

\item[-] \textbf{Unknown data set}: Usually clusters have a centroid with the highest $\rho$, and a set of neighbours with smaller $\rho$. For illustration, in the top panel of Fig. \ref{fig:offset_simple} we can observe the SNR of a cluster with this behavior. In a realistic setting, we do not know the GPS time of the centroid of the cluster $t_{cluster}$ \textit{a priori}, so we define them following a procedure similar to GstLAL. We order all the triggers according to their GPS time. {Afterwards, we divide the whole observing run in windows of $\pm 1\,$s, starting from the beginning of the run.} Within each window, we select the trigger with maximum $\rho$ as the centroid, so that $t_{cluster} = t_{\max{\rho}}$. As before, we select $t_{w} = \pm \,\Delta t$ around $t_{cluster}$ to define the cluster.
\end{itemize}

{While the methodology outlined above provides a systematic approach to identifying and analyzing matched-filtering clusters, it is a simplified approach with several limitations. One significant caveat is the assumption that the trigger with the highest SNR within a given window accurately represents the true centroid of a cluster in the \textit{unknown data set}. This approach may lead to inaccuracies if multiple signals or glitches overlap within the window, potentially causing the algorithm to misidentify the centroid or miss other relevant triggers with slightly lower SNR. Additionally, the fixed window size, although optimized for specific signal characteristics like IMBH signals, might not be ideal for unexpected GW signals of unknown length, leading to suboptimal clustering in some cases. These factors could introduce biases or reduce the accuracy of the clustering, highlighting the need for further refinement and validation of the clustering algorithm in future work.}

\subsection{Trigger tracks}
\label{sec:features}

\begin{figure*}[]
\subfloat[\label{fig:imbh}{Triggers of \texttt{Injection} class}]{%
\includegraphics[width=0.5\textwidth]{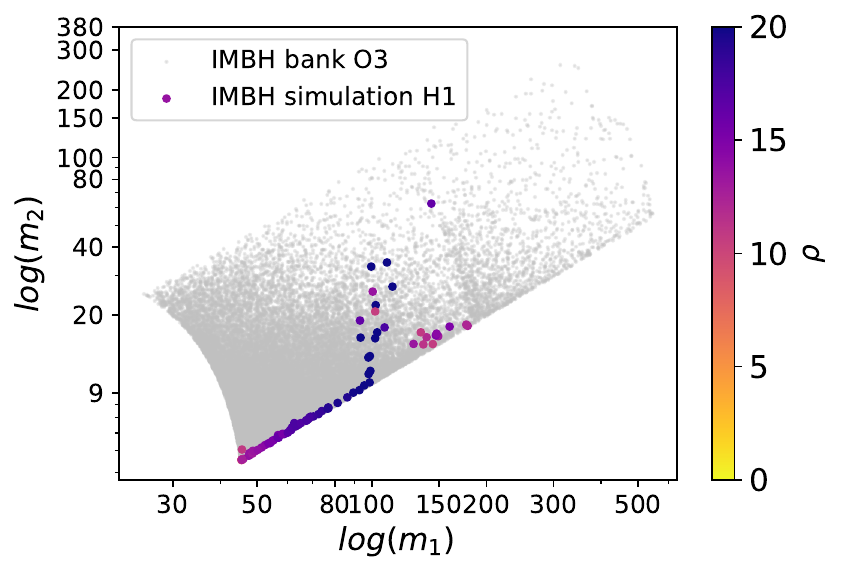}%
}\hfill
\subfloat[\label{fig:blip}{Triggers of \textit{Blip} class}]{%
\includegraphics[width=0.5\textwidth]{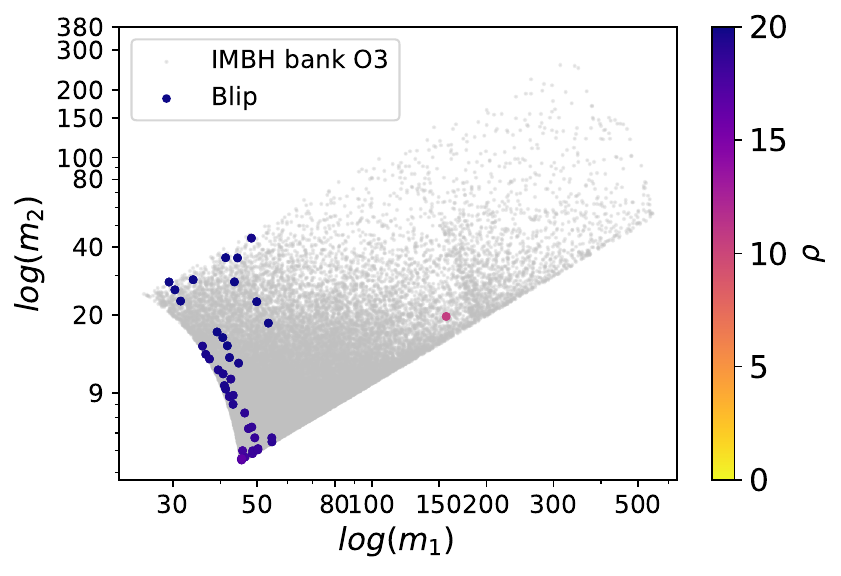}%
}\hfill
\caption{O3 IMBH template bank as a function of the progenitor masses in logarithmic scale, where every grey point represents a template from the bank. Triggers associated with an \texttt{Injection} or a  \textit{Blip} glitch are coloured according to the maximum SNR, $\rho$. }
\label{fig:tracks}
\end{figure*}

 As  we mentioned in Section \ref{sec:triggers} when  the SNR associated with a template is over a certain threshold, it will produce a trigger with intrinsic parameters $\lambda_{int}$ associated. In this particular work, we reduce the single-detector $\lambda_{int}$ associated with the $i$th template forming the feature vector $\phi_{i}$,

\begin{equation}
\label{eq:features}
\phi_{i} = \{\rho_{i}, \xi_{i}, m_{1, i},  m_{2, i},  s_{1z, i},  s_{1z, i} \},
\end{equation}

containing the SNR $\rho$, the consistency check $\xi$ (see Eq. \ref{eq:chisq}), the masses of the progenitors ($m_{1}, m_{2}$) and the $z$-component of their spins ($\xi_{1}$, $\xi_{2}$), respectively. As we mentioned before, {an unknown signal} $s(t)$ might have {an  associated \textit{track}}, so to illustrate this scenario we show in Fig. \ref{fig:tracks} the {tracks} generated by an \texttt{Injection} (Fig. \ref{fig:imbh}) and a \textit{Blip} glitch (Fig. \ref{fig:blip}) {projected in the mass and SNR space}. We show the O3 IMBH template bank as a function of the progenitor masses $m_{1}$ and $m_{2}$, where every dot represents a template. We also colour the tracks, where their colour  is related to the maximum SNR, $\rho$. As we can see in Fig. \ref{fig:imbh}, the \texttt{Injection} matches {a concrete space in the} low-mass {region} at $\rho \sim 10$, {and a slightly sparse space in the higher-mass region} at $\rho \sim 20$. On the other hand, in Fig. \ref{fig:blip} the templates matching the \textit{Blip} glitch are at $\rho \sim 12$, being sparse in the progenitor mass dimension. {Assuming that the distribution of GW and glitch tracks possess distinct underlying features, we can employ ML methods to differentiate them.}

\subsection{Time shift and false alarm rate}\label{sec:timeshift}

{Our model will compute a certain statistic $\Lambda^*$ for the tracks obtained from matched filtering the detector data. We refer to this resulting distribution as \textit{foreground}. 
It is challenging to assess if single-detector tracks have astronomical or terrestrial origin, but time-coincident signals in multiple detectors thousands of kilometers apart are more likely to be astronomical.  

To assess the significance of our foreground, we need to compare it with an noise-only distribution, referred to as \textit{background}.}
The background tracks time-coincident is multiple detectors must be uncorrelated, so we can use the time-sliding technique, which time-shifts the detectors with respect each other. Subsequently, the coincident triggers observed in the time-shifted data will serve as the background triggers, {which are accidental coincidences and not due to actual astrophysical events}. 

{To increase the significance of our foreground trigger, we need to compare it with an extensive population of background triggers. Hence, we make the process of time-sliding the data $\Delta t_{l}$ iterative by defining the time shift $T_{l} \equiv l\Delta t_{l}$, where $l$ is known as \textit{lag}. When $l=0$ we recover the foreground, which is also {known} as \textit{zero-lag}. }
{A relevant quantity is the total observing time of the foreground or \textit{search time} $T_{f}$, and the total observing time of the background $T_{b}$. In our setup, for the $l^\text{th}$ time-shift $T_{f} > T_{b, l}$, since the overlap between detectors will decrease with increasing offsets. {Nonetheless, it is possible to perform time shifts on a \textit{ring}, where data that slides past the end of the segment is placed at the beginning. Note that even in this scenario, time shifts will result in different durations due to gaps in the data.} 
To achieve a significant estimation of the background we would perform $L$ time shifts, such that $T_{f} \ll T_{b} = \sum_{l=1}^{L} T_{b, l}$.}

{To assess the significance of our track with statistic $\Lambda^{*}$ from the foreground, we compare it with tracks produced within the background, known as \textit{false alarms}.  We can define the \textit{false alarm probability} $P_{F}$, as the probability of obtaining $j \geq 1$ triggers with a statistic $\Lambda \geq \Lambda^{*} $ within the background, during the time of the search $T_{f}$,}

 \begin{equation}
 \label{eq:falsealarmprobability}
 P_{F}(j \geq 1| \mathcal{F}(\Lambda^{*}), T_{f}) = 1- e^{-\mathcal{F}(\Lambda^{*})T_{f}},
 \end{equation}
{where $\mathcal{F}(\Lambda^{*})$ is called the \textit{false alarm rate} (FAR) of a trigger with statistic $\Lambda^{*}$. For our purposes, we will compute the FAR using the following estimator:}
%{\noindent Here, we define $\mathcal{F}(\Lambda^{*})$ as the \textit{false alarm rate} of a trigger with statistic $\Lambda^{*}$, measured as}
\begin{equation}
\widetilde{\mathcal{F}}(\Lambda^{*}) = \frac{\sum_{l=i}^{L}\widehat{N}_{l}(\Lambda \geq \Lambda^{*})}{T_{b}},
\end{equation}
{where $\widehat{N}_{l}$ represents the false alarms of the $l^\text{th}$ time-shift, i.e.  the number of background triggers within the $l^\text{th}$ time-shift with statistic $\Lambda \geq \Lambda^{*}$. As before, $T_b$ is the total background time. It is relevant to note that for each slide the duration of $T_{b,l}$ will vary, so we define the \textit{effective number of slides} $\widetilde{N} \equiv T_{b}/T_{f}$, so that}

\begin{equation}
\begin{split}
& \widetilde{\mathcal{F}}(\Lambda^{*}) = \frac{\sum_{l=i}^{L}\widehat{N}_{l}(\Lambda \geq \Lambda^{*})}{T_{b} T_{f}/T_{f}} \\
&= \frac{\widehat{N}(\Lambda \geq \Lambda^{*})}{T_{f} \widetilde{N}} = \frac{\overline{N}(\Lambda \geq \Lambda^{*})}{T_{f}}.
\label{eq:far}
\end{split}
\end{equation}

{We define $\widehat{N}(\Lambda \geq \Lambda^{*})$ as the total number of background triggers with $\Lambda \geq \Lambda^{*}$ in all time shifts, while $\overline{N}(\Lambda \geq \Lambda^{*})$ is the average number of false alarms in a single experiment with a $T_{f}$ duration.}

 Since the oﬀsets of the time shifts should be large enough that each slide can be thought of as an independent experiment, we chose to slide L1 in steps of $3\,$s, and V1 in steps of $6\,$s with respect to H1, which is fixed. We performed enough time shifts to produce $1,000\,$yr of data in triple detector coincidence. 
Afterwards, the time shifts are binned according to their coincidence type and their time of background.

To minimize the number of false positives, the types of detector time coincidences analyzed in this work are either double or triple. Using Eq. \ref{eq:far}, it is necessary to properly measure the time of the search $T_{s}$ and the time of the background $T_{b}$. If we are measuring the time in double-time coincidence, for example, between H1 and L1, it would imply that V1 was down at that time (no V1). Otherwise, if all detectors record data simultaneously, it is considered a triple-time coincidence. In Table \ref{tab:background_time} we can see an overview of the time of the search $T_{s}$. Since H1L1 (no V1) is extremely small, it is computationally intensive to generate $1000\,$yr of time shifts. { Thus, in the interest of time and computational resources we focus on triple-coincident time as we expect a better performance.}

As in a standard search, to assess if two tracks are coincident, we only consider tracks within the time coincidence window. For example, using the H1L1V1 time-coincidence, we can have tracks that occur in all detectors simultaneously and tracks that occur only in H1 and L1. Thus, we will have double coincident tracks, such as H1L1, during H1L1V1 time, as well as triple coincident tracks in H1L1V1 during H1L1V1 time.

\vspace{-2mm}
\begin{table}
\caption{Total time of search ($T_{s}$) for different detector combinations, measured in years.}\label{tab:background_time}
\begin{tabular}{@{}cccc@{}}
\toprule
\textbf{} & \textbf{Time of Search ($T_{s}$)} \\ \midrule
\textbf{H1L1 (no V1)} & 0.0000382 \\
\textbf{H1V1 (no L1)} & 0.1117765 \\
\textbf{L1V1 (no H1)} & 0.1338800 \\
\textbf{H1L1V1}       & 0.5619618 \\ \bottomrule
\end{tabular}
\end{table}

\section{Methodology}
\label{sec:methodology}

The challenges in GW research require innovative solutions, and ML has emerged as a crucial tool for addressing them due to its adaptability and transversality. The main advantage of ML techniques is their rapidity {during the deployment phase}, since most of the computations are made during the training stage. 
In the past few years, researchers have explored different ML applications to GW data analysis, such as detection of CBC \cite{George:2016hay, PhysRevD.105.043006, ruder2016overview, Gabbard:2017lja, Sharma:2022ibm, Krastev, Menendez-Vazquez:2020khz, Baltus:2021nme, Baltus:2022pep, 2024arXiv240905068M, Nousi:2022dwh, Koloniari:2024kww}, burst identification \cite{2020arXiv200204591C, PhysRevD.105.084054, Iess:2020yqj, Iess:2023quq, LopezPortilla:2020odz, Lopez:2021rci, Meijer:2023yhn, Boudart:2022xib, Boudart:2022apz, Skliris:2020qax}, glitch characterization \cite{Biswas:2013wfa, zevin2017gravity, Glanzer:2022avx, bahaadini2018machine, Laguarta:2023evo} and synthetic data generation \cite{Lopez:2022lkd, Lopez:2022dho, Powell:2022pcg,Yan:2022spw, Dooney:2022arh, Dooney:2024pvt}, among others. We refer the interested reader to \cite{Cuoco:2020ogp} and \cite{Schafer:2022dxv} for a  review.

In \cite{PhysRevD.105.043006} the authors seek to enhance current state-of-the-art search algorithms with ML applications. Following a similar line of reasoning{, and as previously mentioned,} we use the triggers generated by the IMBH search of GstLAL during O3 to train a classification algorithm for multiple classes. In this way, we output a statistic to differentiate between IMBH signals and glitches, providing direct information about the nature of these populations. 

\subsection{Feature vector}
\label{sec:feature_vector}

Under the assumption that different signals have different track structures, it is possible to learn them with ML. Tracks have data structures known in ML as \textit{multi-instance representation}, where $N$ feature vectors $\phi_{i}$, also known as triggers, represent a given example. This implies that each track may contain a different number of triggers for each example. Consequently, the input length to our multivariate ML model changes from one example to another. {This poses a limitation because standard multivariate ML techniques only accept fixed-length inputs.}
{As a proof-of-concept to overcome this obstacle, we average all the feature vectors $\phi_i$ associated with a given signal by weighting them with $\rho$,  resulting in the following input feature}

\begin{equation}
\begin{split}
&\label{eq:feature}
\Phi = \{\bar{\rho}, \bar{\xi}, \bar{m}_{1},  \bar{m}_{2},  \bar{s}_{1z},  \bar{s}_{1z} \} \\ & \text{ where } \bar{x} = \frac{\sum_{i=1}^{N} x_{i}\rho_{i}}{\sum_{i=1}^{N} \rho_{i}},
\end{split}
\end{equation}
for $i \, \in 1, \dots, N$ {the triggers associated with each signal}. In this way, we give more weight to the template that best ``{matches}" a given signal. This weighted average enhanced the performance of our multivariate ML model with respect to a standard average.

\subsection{Tackling class imbalance}
\label{sec:kfold}

\begin{table}
\caption{Original data set size before sampling}\label{tab:original_size}
\begin{tabular}{@{}cccc@{}}
\toprule
                                                    & \textbf{\begin{tabular}[c]{@{}c@{}}Hanford\\ (H1)\end{tabular}} & \textbf{\begin{tabular}[c]{@{}c@{}}Livingston\\ (L1)\end{tabular}} & \textbf{\begin{tabular}[c]{@{}c@{}}Virgo\\ (V1)\end{tabular}} \\ \midrule
\multicolumn{1}{c|}{\texttt{Injections}}            & 85107                                                           & 101307                                                             & 54436                                                         \\
\multicolumn{1}{c|}{\texttt{Blip}}                  & 2717                                                            & 1701                                                               & 1534                                                          \\
\multicolumn{1}{c|}{\texttt{Fast\_Scattering}}      & 114                                                             & 18589                                                              & -                                                             \\
\multicolumn{1}{c|}{\texttt{Koi\_Fish}}             & 5147                                                            & 4061                                                               & 731                                                           \\
\multicolumn{1}{c|}{\texttt{Low\_Frequency\_Burst}} & 1523                                                            & 109                                                                & 3044                                                          \\
\multicolumn{1}{c|}{\texttt{Tomte}}                 & 537                                                             & 18619                                                              & 674                                                           \\
\multicolumn{1}{c|}{\texttt{Whistle}}               & 1946                                                            & 95                                                                 & 246                                                           \\ \bottomrule
\end{tabular}
\end{table}

{While we can simulate a large number of GW signals, glitches have a finite nature. Despite their occurrence rate being approximately $\sim 1 \,\text{min}^{-1}$, some classes of glitches are more common than others~\cite{LIGOScientific:2020ibl}.} 

{For example, in L1 the \mlp{Gravity Spy pipeline} identified $> 10^4$ \texttt{Tomtes}, but only $\sim 10^3$ \texttt{Whistles}}. {In Table \ref{tab:original_size} we show the original size of our training and validation set, which is highly imbalanced}. This is problematic for classification ML algorithms, as heavily imbalanced datasets tend to be biased towards the majority class to minimize their loss function. This may wrongly be interpreted as a ``good" performance, while in reality, the model is only learning a single class: the largest one.
Similarly to \cite{Powell:2022pcg}, to circumvent this issue we use undersampling or oversampling techniques:

\begin{itemize}
\item[-] \textbf{Undersampling:} In our data set, the number of \texttt{Injections}, i.e. simulated IMBH signals, performed is significantly higher than the largest glitch class. Thus, we randomly sample this class to match the size of the oversampled glitches.
\item[-] \textbf{Oversampling:} To oversample we use bootstrapping with replacement, which stochastically resamples the existing data set allowing the same example to appear more than once \cite{James2013}. {Using this method, we oversample all glitch classes to equal the size of the largest glitch class.}
\end{itemize}

While this method improved the performance of our ML method,  it is important to realize that bootstrapping with replacement does not generate new data nor gives new information about the classes. If the original data set does not represent the class population, it will bias the ML algorithm producing overfitting.

\begin{figure}
    \centering    \includegraphics[width=0.45\textwidth]{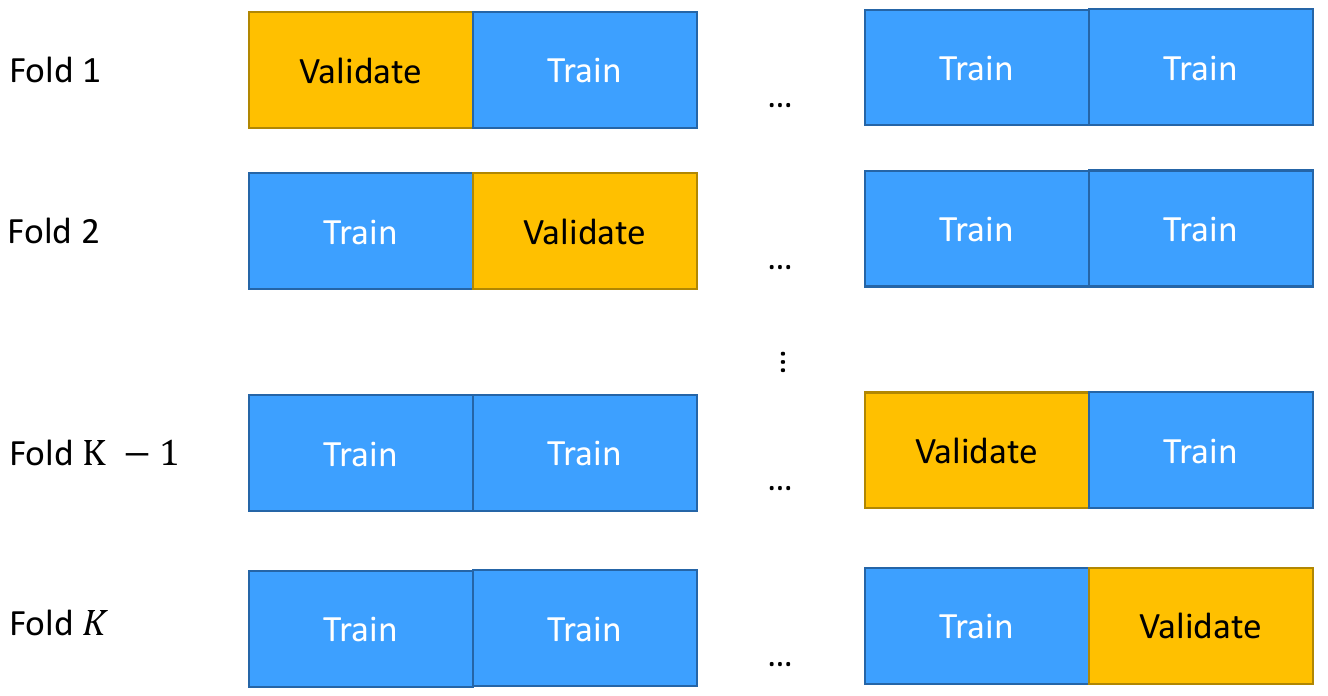}
    \caption{Illustration of $K$-fold cross-validation.}
    \label{fig:kfold}
\end{figure}

To avoid overfitting we use \textit{K-fold cross-validation} \cite{journals/jmlr/BengioG04}. This method partitions the data set in $K$ subsets or folds (see Fig. \ref{fig:kfold}). The model is evaluated $K$ times, using a different fold as validation in each iteration, while training on the rest of the folds.
 This allows for a more robust estimate of the model’s performance, as it accounts for the variability in the data and limits the potential for overfitting. In particular, we chose  $K=9$ as it is a trade-off between computational complexity and performance.

As we mentioned in Section \ref{sec:clustering} we want to test the generalization ability of our ML method for O3a and O3b. Therefore, we use the \textit{known data set} of O3a for the standard ML procedure of training, validation and testing, splitting the data set into $80\%$ for training, $10\%$ for validation and $10\%$ for testing with the $K$-fold cross-validation method. Once we have tested in \textit{known data set}  of O3a, we use the whole \textit{known data set} of O3b to test the generalization ability of our method.% Last but not least, we {explore} the whole \textit{unknown data set} to find real GW signals (see results in Section \ref{sec:unknown}).

\subsection{Training model}
\label{sec:mlp}

In this work, we use a multi-layer perceptron (MLP), also known as a fully connected or dense network. This model implements a mapping between the input space $\mathds{R}^n$ to an output space $\mathds{R}^d$, $f_{\theta}: \mathds{R}^n \rightarrow \mathds{R}^d$, where $\theta$ represents its trainable parameters.

The unit of an MLP is the perceptron, a mathematical simplification of human neurons \cite{rosenblatt1958perceptron}. In this unit, the information is given by a $n$-vector $x = [x_1, \dots, x_n]$, that travels in the neuron and interacts multiplicatively with the weights, $w = [w_1, \dots, w_n]$. The goal of the perceptron is to learn the value of the weights, which controls the strength of the influence of the previous neuron. In this simplistic model, the information is summed, such that if the sum is above a certain threshold, the neuron creates or \textit{fires} a signal. This \textit{firing rate} is modelled as the activation function $f: \mathbb{R} \rightarrow \mathbb{R}$, which is non-linear. Thus, we can mathematically define the $a$ output, known as activation as,

\begin{equation}
\label{eq:perceptron}
a = f\left(\sum_{i}^{n}w_{i}x_{i} + b\right),
\end{equation}
where $b$ is the bias term, which can be interpreted as the error of the model. The MLP model is a stacking of neurons, where neurons are arranged in layers. In this setup, for a given layer $l$ we can define the matrix of weights $W_{ij}^{(l)}$ and the vector of biases $b_{i}^{l}$, where the weights are the connection between the unit $j$ of layer $l$ and the unit $i$ of layer $l+1$, and the bias is associated with the unit $i$ of layer $l+1$. Then, we can express the MLP compactly as

\begin{equation}
\begin{aligned}
&z^{(l+1)} = W^{(l)}a^{(l)} + b^{(l)}, \ \ a^{(l+1)} = f(z^{(l+1)}),\\
& \text{where } z_{i}^{l+1} = \sum_{j=1}^{2} W_{ij}^{(1)}x_{j} + b_{l}^{l}.
\end{aligned}
\end{equation}

To differentiate IMBH signals from glitch classes (see Sections \ref{sec:glitches}), we construct an MLP (discussed in Section \ref{sec:mlp}) for each detector, that inputs the feature vector $\Phi$ (see  Eq. \ref{eq:feature}) and outputs a probability vector that indicates the likelihood for each class. The size of this probability vector depends on the detector: we distinguish 7 classes for H1 and L1, but 6 classes for V1, as \texttt{Fast\_Scattering} class is not present in that detector.

We implemented the MLP structure in \textit{PyTorch} \cite{NEURIPS2019_9015}. After several experiments, we selected the best-performing architecture from our results: it consists of 3 hidden layers of 350 each, using the {ReLU activation function \cite{relu}}, except the output layer that uses Softmax activation function, {suitable for multi-class classification,} since it is a multi-class task. This MLP architecture is common for each GW detector, but they have been trained separately, i.e. it is a single-detector classifier. To optimize the networks we use cross-entropy loss function, and Adam optimizer \cite{Kingma:2014vow}. Moreover, to adjust the number of epochs and avoid overfitting we implemented an \textit{early stopping} algorithm \cite{Yao2007}. We define an epoch as the number of times the network has passed through the whole training and validation data set. Thus, the early stopping algorithm calculates the difference in  validation accuracy $\mathcal{A}_{val}$ between the current epoch,  $e$, and the best epoch, $e^{\text{best}}$. The training process finishes if
\begin{equation}
|\mathcal{A}^{e}_{val} - \mathcal{A}^{e^{\text{best}}}_{val}| \leq \epsilon, 
\label{eq:earlystopping}
\end{equation}
during 150 epochs, with $\epsilon = 0.0001$. 
For the learning rate, we use an \textit{adaptive learning rate} built-in the \textit{PyTorch} function \texttt{ReduceLROnPlateau}. Setting the initial learning rate  to $10^{-3}$, if the validation accuracy remains constant after $100\,$epochs, the learning rate decreases $10\%$. The combination of these methods proved to increase the performance while decreasing the time needed for fine-tuning.

\subsection{Time coincident tracks}
\label{sec:time_coincident}

Even though the MLP receives direct information from the \texttt{Injection} class and various glitch classes, misclassifications are still common due to the simplicity of the feature vector (see Section \ref{sec:feature_vector}). Nonetheless, if a potential GW signal is detected in multiple detectors, the likelihood of it being of astronomical origin increases significantly. Hence, it is standard in GW searches to evaluate triggers that happened within the light time travel between detectors: $10\,$ms between H1 and L1, $27\,$ms between H1 and V1, and $26\,$ms between L1 and V1. An additional $5\,$ms  is considered for statistical fluctuations  \cite{Sachdev:2019vvd}.

For the \texttt{Injection} class of the \textit{known data set}, we can identify coincident tracks because we have access to the ground truth. However, for the glitches of the \textit{known data set} and the \textit{unknown data set} itself, this information is not available \textit{a priori}, so we need to define criteria for when two tracks are time coincident. In a standard matched-filtering search, a time coincident trigger occurs when the same template triggers in different detectors within the light travel time between them. 

In this work, the approach is inherently different, as we do not deal with individual triggers but with averaged tracks. Thus, two tracks from different detectors are considered time coincident if the average time of the triggers within each track falls within the light travel time and time fluctuations.
Nevertheless, even if two tracks are time coincident, there is no guarantee that they originate from the same signal. Therefore, two tracks are only considered time coincident if they have at least one common trigger. If tracks from different  detectors coincide in time, we assign to the coincidence a probability equal to the harmonic mean, one of Pythagorean means 
\cite{538b1c7d-3917-388b-be54-a65758913e9f, harmonic}, of each trigger, relabelling our statistic for $ N$ detectors as
\begin{equation}
\label{eq:harmonic_mean}
\bar{P}_{inj} = N/(\sum^N_{k} P^{-1}_{inj, k})
\end{equation}
where $P_{inj}$ is the probability of being an \texttt{Injection} (see Section \ref{sec:glitches} for a description). It is important to note that the MLP performs a single-detector inference, and the time coincident step is computed independently afterwards. In future works, it would be interesting to provide the full information of the time coincident tracks to a ML algorithm. This integration could enhance the model's ability to distinguish GW signals from glitches.

\section{Results}
\label{sec:results}

In the present Section, we show the results of the performance of the MLP model with H1 data. The results for L1 and V1 can be found in the Appendix \ref{sec:acknowledgements}. Firstly, in Section \ref{sec:tw}, we describe the selection of the time window $t_{w}$ for the \textit{known data set} of O3a{, i.e. the controlled data set of O3a composed of simulated GW signals and well-known glitches}. Afterwards, in Section \ref{sec:train}, we show the results of the training and validation with the previous data set. In Section \ref{sec:performance} we test the model with the \textit{known data set} of O3a and O3b. In Section \ref{sec:performance_gw} we assess the significance of the ML inference with the accidental background of time shifts. %Finally, in Section \ref{sec:unknown} we explore \textit{unknown data set} of O3, searching for real GW signals.

\subsection{Selecting a time window}
\label{sec:tw}

As mentioned in Section \ref{sec:clustering}, we explored different time windows $t_{w}$ for each GW detector, namely H1, L1 and V1. To evaluate the performance of the  algorithms without fine-tuning them we use the receiver operator characteristic (ROC) curve,  which is represented as the true positive rate (TPR)  as a function of false positive rate (FPR) \cite{reference/ml/Ting10}. 
A relevant point is that our task is a multi-class classification, so to compute the ROC curve, which is usually employed in binary classification, we need to reduce our problem to a pairwise comparison, i.e. \texttt{Injection} class  (positive class) against all other classes (negative class). 

%\vspace{3mm}
\begin{figure}[h]
\subfloat[\label{fig:twh1o3a}{ Testing on O3a with \textit{known data set}}.]{%
\includegraphics[width=0.5\textwidth]{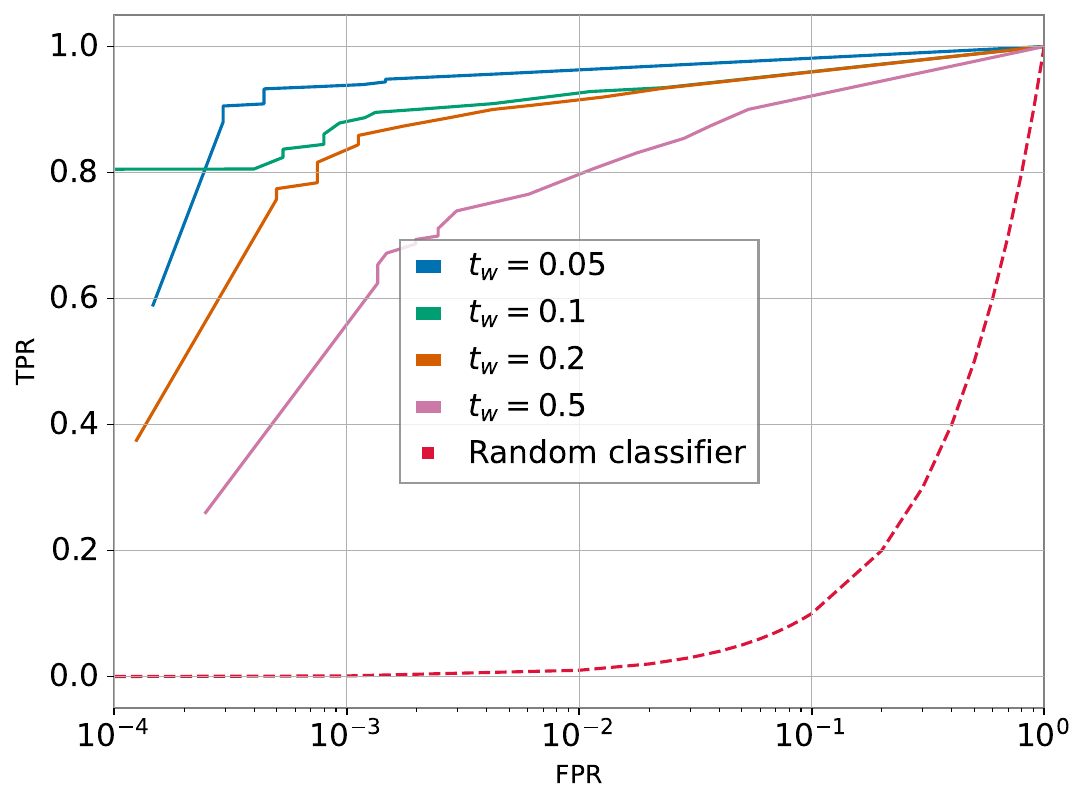}%
}\hfill
\subfloat[\label{fig:twh1o3b}{Testing on O3b with \textit{known data set}}.]{%
\includegraphics[width=0.5\textwidth]{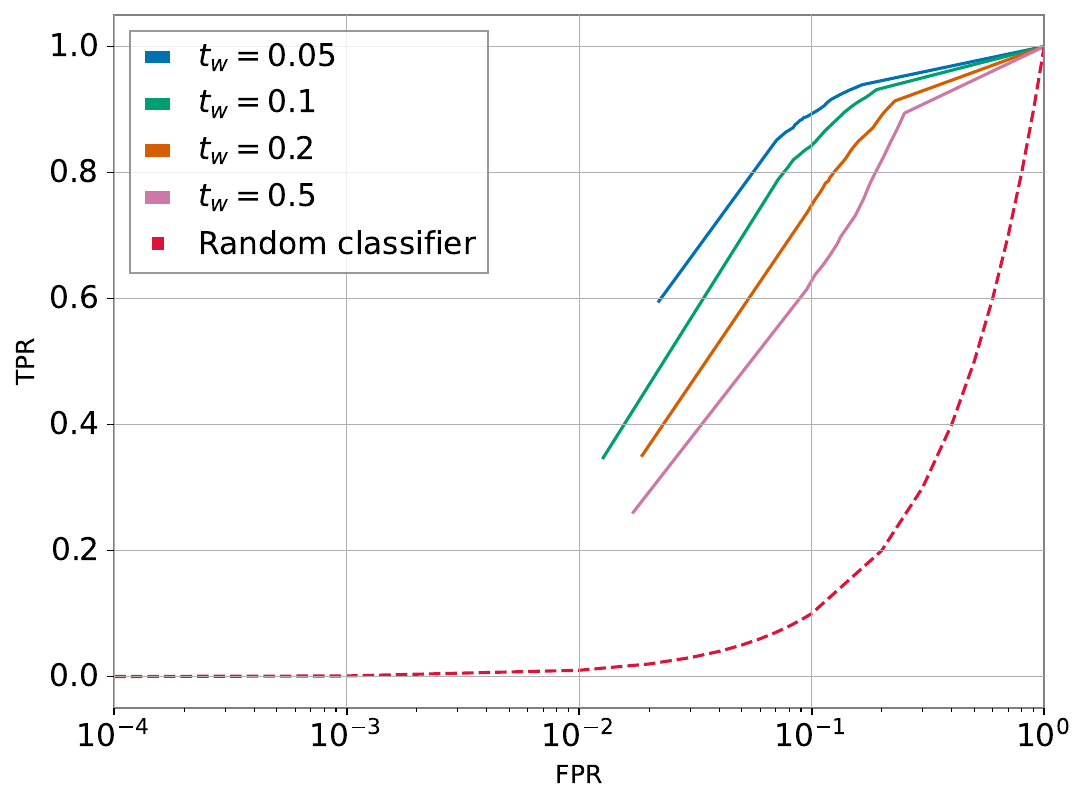}%
}\hfill
\caption{Receiver operator characteristic (ROC) curve for different time windows $t_{w}$ in H1, i.e. true positive rate (TPR) as a function of false positive rate (FPR) in logarithmic scale. The positive class is \texttt{Injection} class, while the negative class is any of the other classes. \textit{(Top)} Testing in the \textit{known data set} of O3a. \textit{(Bottom)} Testing in the \textit{known data set} of O3b. Note that the dashed line indicates a random guess.}
\label{fig:tunetwH1}
\end{figure}

Furthermore, it is custom in ML to calculate the area under the ROC curve as an evaluation metric, since models with a larger area under the curve have a better performance. However, in the field of GW, it is required to know the performance of the model at low FPR  as we want to minimize the number of FPs when claiming detection, i.e. the number of glitches incorrectly classified as GW signals. For this aim, we select a grid of decision thresholds $\theta^*$ in range $[0.1, 0.9]$ with a spacing of $0.1$ and in range $[0.9, 1.0]$  with a spacing of $0.01$. When a given probability exceeds this value, we classify the input as a positive, i.e. \texttt{Injection} class, otherwise, it is classified as negative, i.e. any other class. As this grid of decision threshold is not usual in ML, the area under the curve has a different magnitude, so instead we select the $t_{w}$ with the best performance in \textit{known data set} of O3a and O3b.

\begin{figure}
    \centering
    \includegraphics[width=0.5\textwidth]{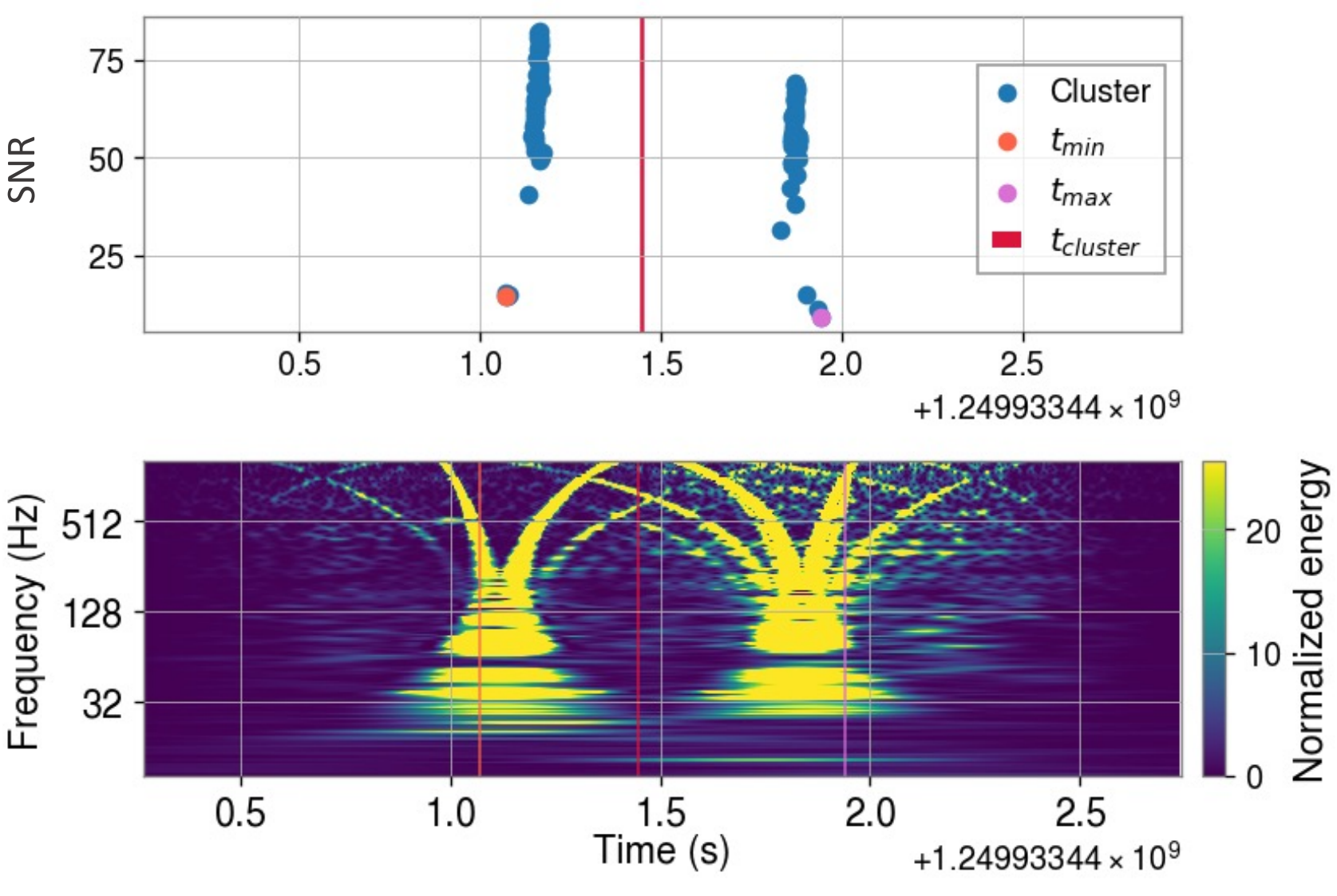}
    \caption{ A \texttt{Whistle} glitch labelled by \textit{Gravity Spy}. \textit{(Top)} SNR of GstLAL tracks as a function of time, where each point represents a template. $t_{min}$ (orange) and $t_{max}$ (pink) mark the beginning and end of the cluster, respectively, while $t_{cluster}$ (red) indicates its centroid given by \textit{Gravity Spy}. \textit{(Bottom)} Q-transform of the time series containing the \texttt{Whistle} labelled by \textit{Gravity Spy}. While \textit{Gravity Spy} labels a single glitch, GstLAL identifies both.}
    \label{fig:offset}
\end{figure}

In Fig. \ref{fig:tunetwH1} we present the ROC curves of H1 for the \textit{known data set} of O3a (Fig. \ref{fig:twh1o3a}) and the \textit{known data set} of O3b (Fig. \ref{fig:twh1o3b}), while the results for L1 and V1 are presented in Appendix \ref{sec:appendix}. In Fig. \ref{fig:twh1o3a} we can observe that the TPR  degrades as we increase the size of $t_{w}$. In  Fig. \ref{fig:twh1o3b} we can see a similar behavior but a higher FPR  range.

%However, at FPR  $< 10^{-1}$ the performance drops dramatically below the random classifier line (dashed line), which implies that the model is making random guesses. 
As we increase the size of $t_{w}$, the performance of the algorithm decreases. This behavior could be caused by the short duration of IMBH signals, meaning that if we define $t_{w}$ to be large, the clusters will contain random triggers that will bias the classification task. Furthermore, this behavior is also observed in L1  and V1 (see Appendix \ref{sec:twl1v1} for details), so we conclude that $t_{w} = 0.05 \,$s is the best time window for our task.

{It is relevant to note that one limitation of this method is the lack of ground truth when defining the centroid $t_{cluster}$ of a glitch. As an example,  in Fig. \ref{fig:offset} we show a \texttt{Whistle} labelled by \textit{Gravity Spy}. The time centre of the cluster, $t_{cluster}$, is marked in red, and according to \textit{Gravity Spy}, it is associated with a single \texttt{Whistle}. GstLAL identifies two clusters that correspond to two \texttt{Whistle} instead, as can be seen from the bottom panel. In future works, it would be relevant to study the effect of this offset on the behavior of the model. }

\subsection{Training with \textit{known data set} of O3a}\label{sec:train}

\begin{figure}[]
    \centering
    \includegraphics[width=0.5\textwidth]{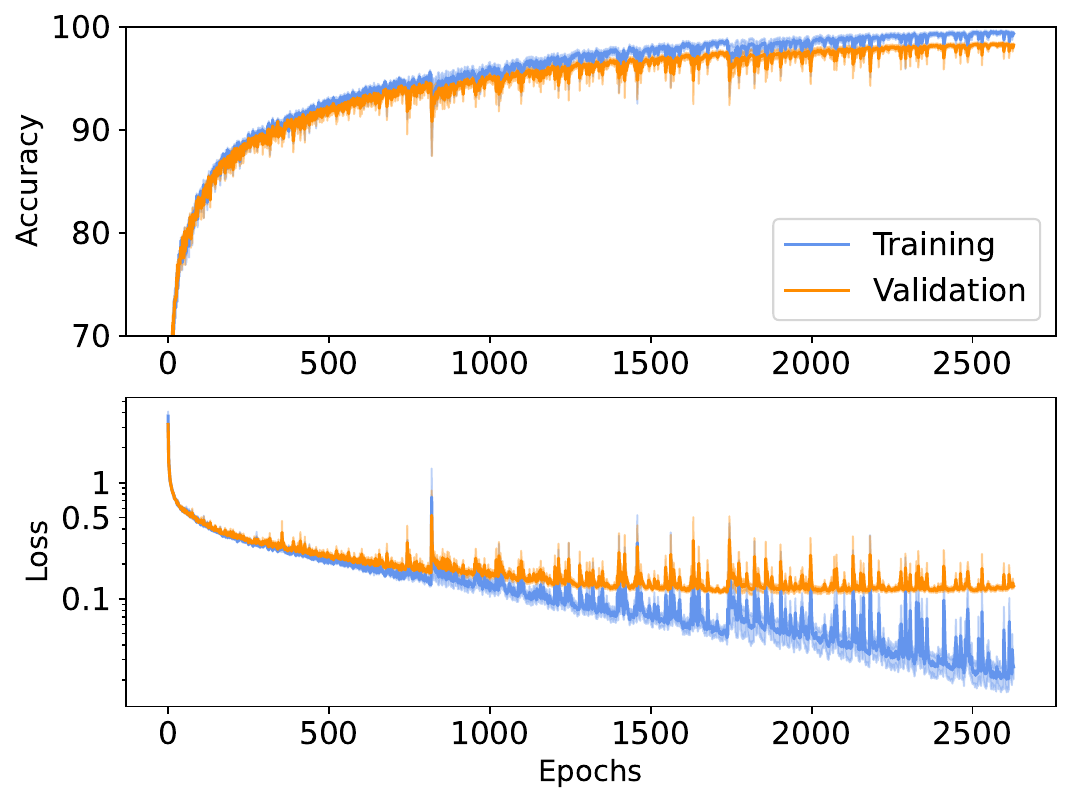}
    \caption{Comparison between training and validating with 9-fold cross-validation for H1 during O3a. \textit{(Top)} Mean accuracy at 3 standard deviations (shaded region) as a function of the epochs during training and testing. \textit{(Bottom)} Average 9-fold cross-validation loss as a function of the epochs during training and testing.}
    \label{fig:losses_H1}
\end{figure}

\begin{figure*}[!htp]
\subfloat[\label{fig:cfmh1o3a}{ O3a \textit{known data set}.}]{%
\includegraphics[width=0.49\textwidth]{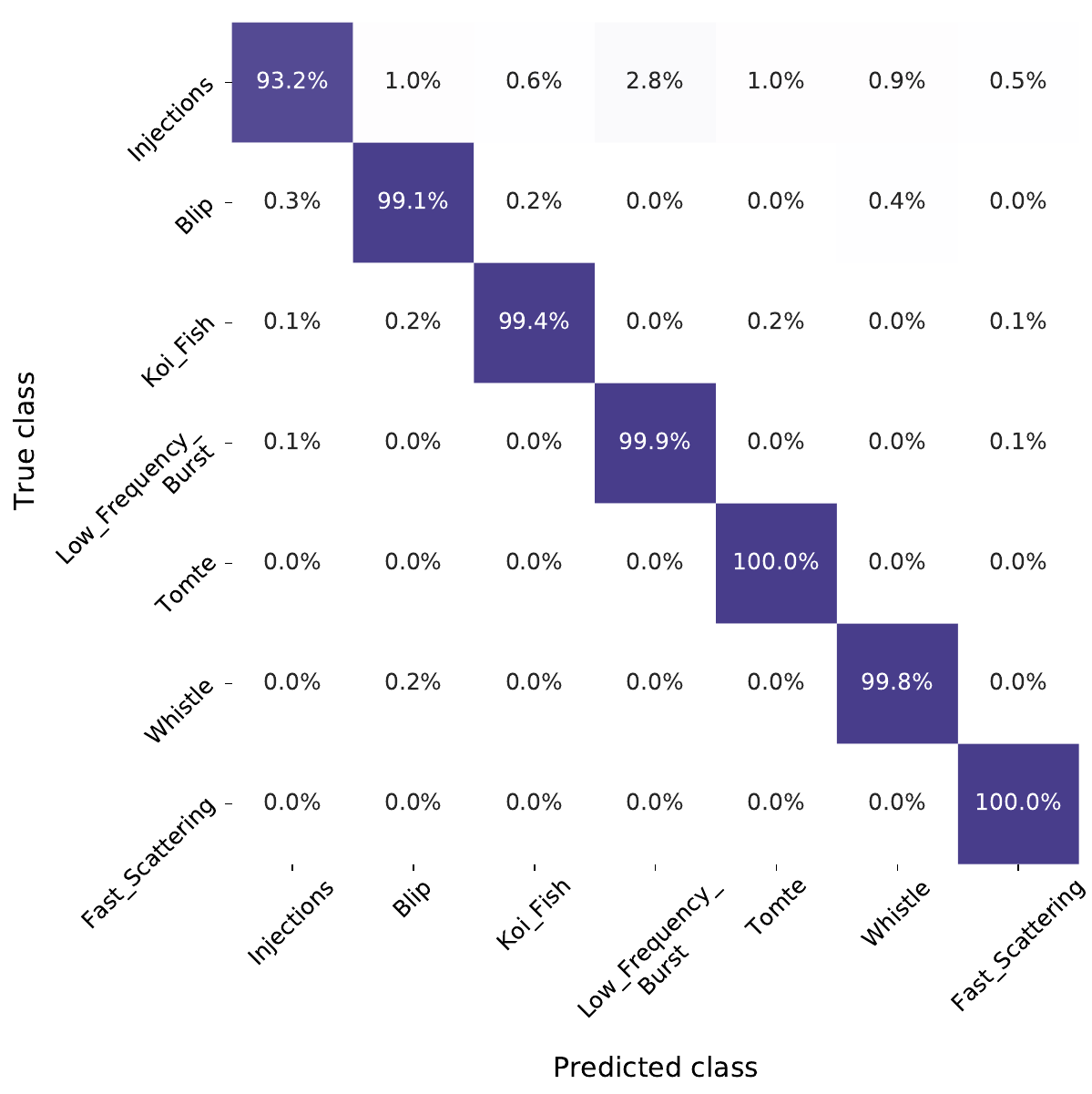}%
}\hfill
\subfloat[\label{fig:cfmh1o3b}{O3b \textit{known data set}}. ]{%
\includegraphics[width=0.45\textwidth]{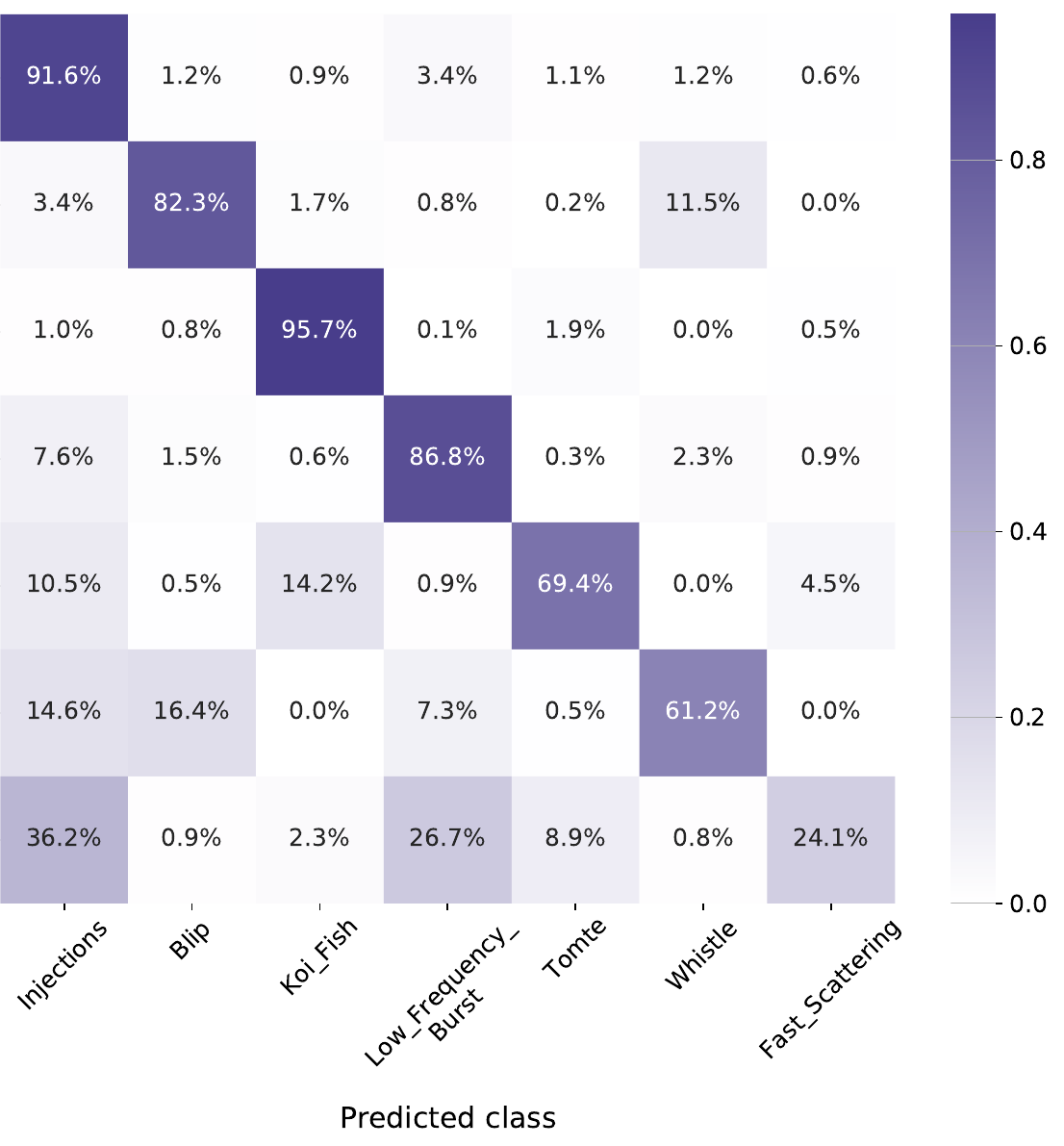}%
}\hfill
\caption{Relative values of the confusion matrix for the test set for H1. The y-axis represents the groud-truth, i.e. the true class, and the x-axis represents the prediction of the model, i.e. the predicted class.}
\label{fig:cfmh1}
\end{figure*}
Once we have selected $t_{w}$, and after several fine-tuning experiments, we can train and validate our model with the \textit{known data set} of O3a. As we mentioned in Section \ref{sec:kfold}, we use 9-fold cross-validation to enhance the performance of the algorithm. Thus, in Fig. \ref{fig:losses_H1} we show, for training and validation, the mean accuracy (top panel) and loss (bottom panel) of the 9-fold cross-validation as a function of the epochs, where the shaded region represents $\pm 3$ standard deviations.
Note that the loss is plotted in logarithmic scale so we can appreciate the difference between training and validation loss is minimal, which implies that the model is not overfitting. Another relevant point is that both mean accuracy and loss seem to have sharp peaks around epochs 900 and 1700, which could imply that the learning is unstable. However, this is a known effect of the adaptive learning rate that we described in Section \ref{sec:mlp}. The described behavior is also present in L1 and V1, whose results can be found in Appendix \ref{sec:trainl1v1}.

\subsection{Diving into the \textit{known data}: machine learning performance}\label{sec:performance}

Employing GPU Tesla V100 with a memory of $16\,$Gb allowed us to train our model in $\approx 17.40\,$h for H1 data, $\approx 15.55\,$h for L1 data, $\approx 31.20\,$h for V1 data. Such a large amount of time is mainly due to the 9-fold cross-validation procedure. Nonetheless, once the models are trained we can predict a single input in $2.9\times10^{-6} \,$s. 
As we also want to test the generalization power of the model between O3a and O3b \textit{known data set}, we present their confusion matrix for H1 in Fig. \ref{fig:cfmh1} (see results for L1 and V1 in Appendix \ref{sec:testl1v1}). While in the y-axis we represent the ground truth, in the x-axis we represent the predictions of the model. Furthermore, the percentages in the rows sum $100 \%$, and the elements in the diagonal were successfully classified.

In Fig. \ref{fig:cfmh1o3a} we show the confusion matrix of the test using the O3a \textit{known data set} for H1, where the accuracy is $98.8 \%$. We can see that $93.2 \%$ of \texttt{Injections} were correctly classified with little amount of misclassifications. However, for the O3b \textit{known data set} the accuracy decreases sharply to $73.0 \%$. We can see in Fig. \ref{fig:cfmh1o3b} that some of the glitches are wrongly classified as \texttt{Injections} or other classes. In particular, the \texttt{Fast\_Scattering} class seems to be the most problematic, as  $36.2 \%$ of them seem to be classified as \texttt{Injections} and $26.7 \%$ as \texttt{Low\_Frequency\_Burst}. This lack of generalization between O3a and O3b is common to L1 and V1, where the accuracy in O3a is $95.8\%$ and $99.3\%$,
 and the accuracy in O3b decreases to $67.5\%$ and 
$75.9\%$, respectively. The misclassification of \texttt{Fast\_Scattering} is also present in L1, but the misclassification of \texttt{Low\_Frequency\_Burst} as \texttt{Fast\_Scattering} is more acute. Interestingly, in V1, the most problematic is the \texttt{Whistle} class (see Fig. \ref{fig:cfml1} and Fig. \ref{fig:cfmv1} in Appendix \ref{sec:testl1v1}).

\begin{figure}[]
\subfloat[\label{fig:tprs}{True positive rate (TPR)}]{%
\includegraphics[width=0.45\textwidth]{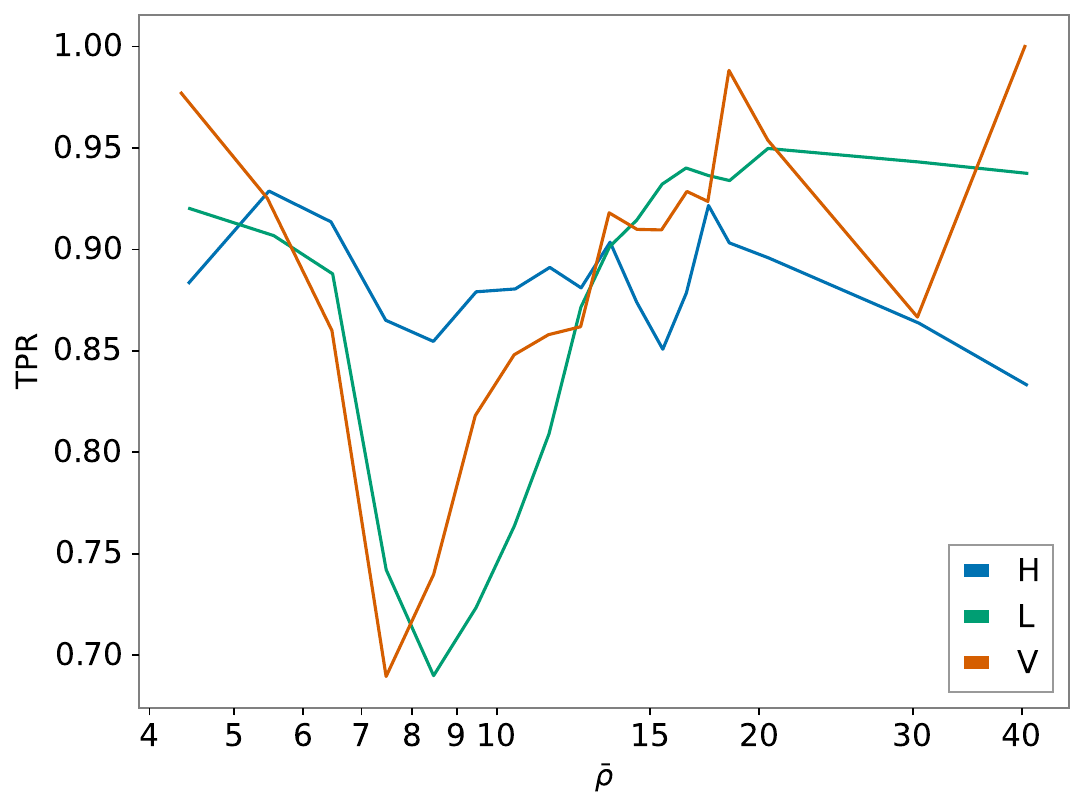}%
}\hfill
\subfloat[\label{fig:snrs}{SNR $\rho$ probability density of different classes. }]{%
\includegraphics[width=0.45\textwidth]{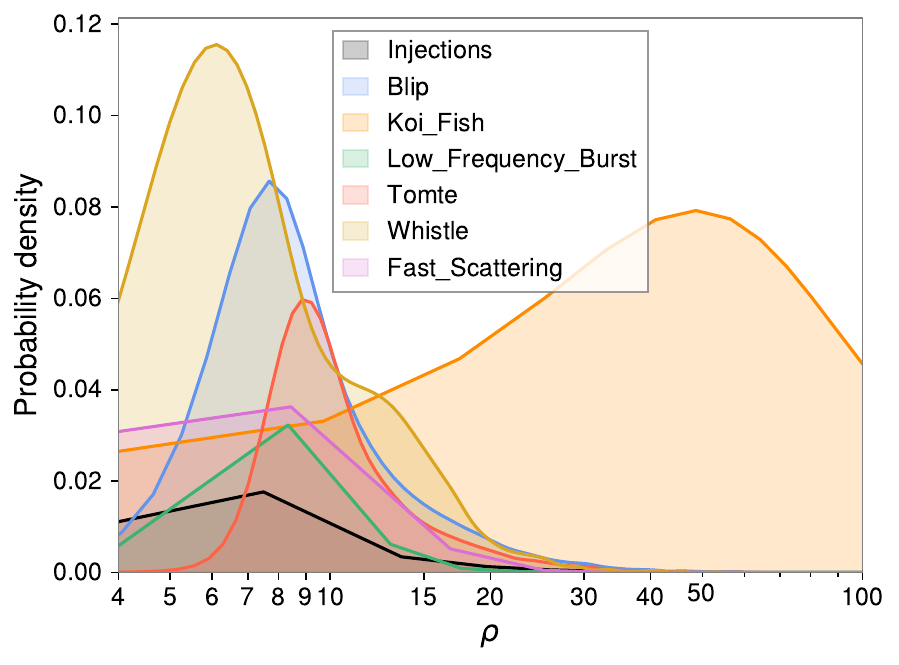}%
}\hfill
\caption{{\textit{(Top panel)} Single detector TPR as a function of the average SNR $\bar{\rho}$ in the \textit{known data set} of O3b for H1, L1 and V1 . \textit{(Bottom panel)}  SNR $\rho$ probability density of different classes in the \textit{known data set} of O3b.}}
\label{fig:tprs_snrs}
\end{figure}

{To assess the degree of misclasification during O3b with respect to the SNR distribution we compute the TPR as a function of the average SNR $\bar{\rho}$ (Fig. \ref{fig:tprs}). For this aim, we arrange the \texttt{Injections} in SNR bins and compute their TPR setting the decision threshold to be $P_{inj}^{*} = 0.9$. Interestingly, we can observe that the TPR is high for low $\bar{\rho}$, showing dip at $ \bar{\rho} \sim 8$ of  TPR$\, \sim 0.85$ for H1 , and TPR$\, \sim 0.7$ for L1  and V1. Moreover, the TPR becomes unstable at $ \bar{\rho} > 15$.}

{To explain this behavior we need to observe the  SNR $\rho$ distributions, computed by \textit{Omicron}~\cite{Robinet:2020lbf}, of the different classes for the \textit{known data set} of O3b (Fig. \ref{fig:snrs}). Firstly, it is important to note that \textit{Gravity Spy} classifies glitches with \textit{Omicron} SNR $> 7.5$ \cite{Glanzer:2022avx}. In Fig.  \ref{fig:snrs} we show the distribution of $\rho$, the resulting SNR from \textit{Omicron}, where we can still observe that most glitch populations are centred around  $ \rho \sim 8$, which explains the dip in Fig. \ref{fig:tprs}. Our method understands that {most classes of problematic glitches, such as \texttt{Fast\_Scattering}, \texttt{Tomte} and \texttt{Low\_Frequency\_Burst}} are around this value, so it does not have trouble distinguishing \texttt{Injections} at lower $ \rho$. {Note that at $\rho < 5$ the model simply differentiates between \texttt{Whistle}, \texttt{Fast\_Scattering} and \texttt{Injection} classes.} Furthermore, as $\rho$ increases, the population of \texttt{Injections} decreases, so due to the low statistic at $ \bar{\rho} > 15$ the TPR seems unstable. {We note that a different SNR distribution for the \texttt{Injection} class might lead to a different behavior of the TPR as a function of the SNR. We will leave this exploration to future work.}}

A possible cause for this lack of generalization between O3a and O3b can be the fact that the interferometers are {evolving} systems, so the glitches produced at the beginning of the observing run can be different from the ones produced at the end {(see Fig. 4 of \cite{LIGOScientific:2021djp})}. Furthermore, we remind the reader that while the \texttt{Injections} class is the actual ground truth, there is some bias in the definition of the glitch classes (e.g. Fig. \ref{fig:offset}). Another reason is the limitation of the model itself, as we are performing a multi-classification task with information on 6 variables (see Eq. \ref{eq:feature}). To lower the number of FPs, i.e. glitches that are incorrectly classified as \texttt{Injections}, in this single-detector task we can use time coincidence between detectors (see Section \ref{sec:time_coincident} for a definition), since it is less common for glitches to appear in several interferometers at the same time. 

\begin{figure}
    \centering
    \includegraphics[width=0.49\textwidth]{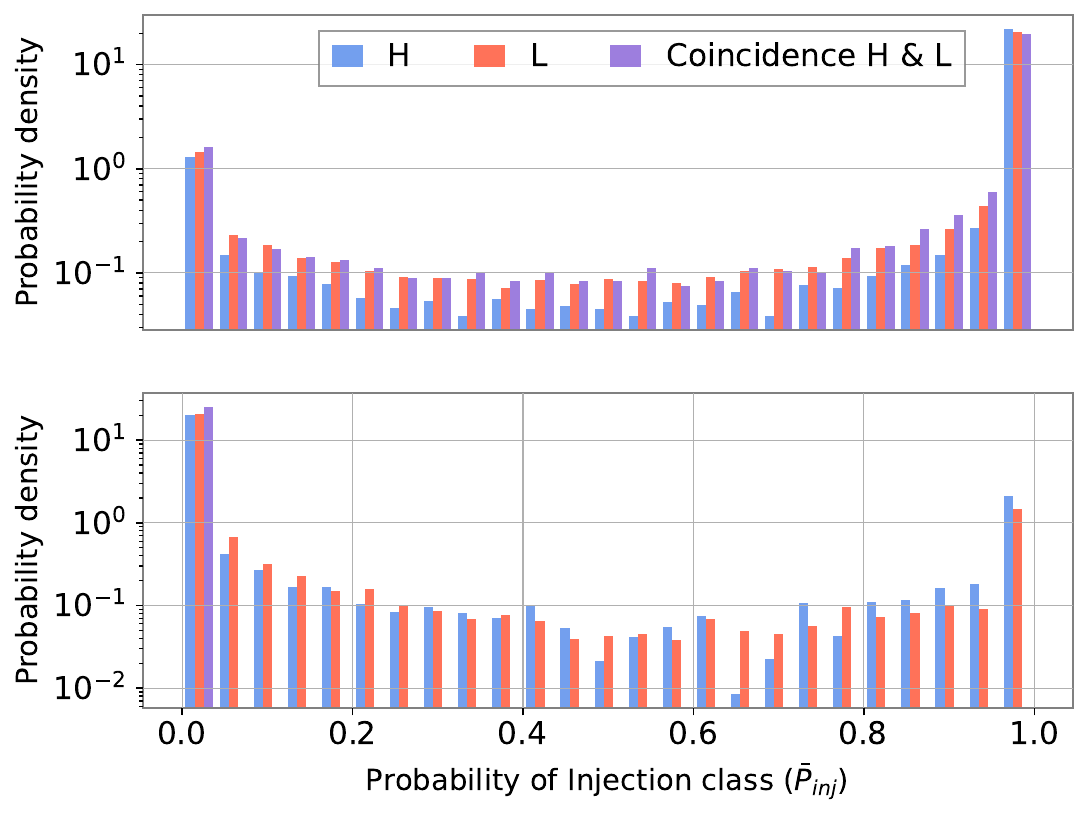}
    \caption{Probability density of being \texttt{Injection}, using the \textit{known} test set of O3b for H1, in blue, L1, in red, and their time coincidence, in purple.   \textit{(Top)} Probability density of elements in the \texttt{Injection} class in logarithmic scale. \textit{(Bottom)} Probability density of elements in all the other classes, i.e. glitches, in logarithmic scale. Given the counts of the $i$th bin $c_i$ and its width $b_i$, we define the probability density as $c_i/(\sum^N_i c_i \times b_i)$, where $N$ is the total number of bins of the histogram.}
    \label{fig:coincidence}
\end{figure}

In Fig \ref{fig:coincidence} we present the probability of being an \texttt{Injection} ($P_{inj}$) for the \textit{known test set} of H1, in blue, L1, in red, and their time coincidence, in purple. In the top panel, we present the probability density of the \texttt{Injection} class, while in the bottom panel, we show the probability density of the glitches. In the top panel, we can observe that, for both detectors, while the MLP model classifies many \texttt{Injections} as such with large $P_{inj}$, there are still many  \texttt{Injections} that have a low $P_{inj}$, which can be misclassified as glitches. Conversely, the MLP model gives many glitches a low $P_{inj}$, but there are still many glitches with a high $P_{inj}$. To increase $P_{inj}$ of \texttt{Injections} and lower the $P_{inj}$ of glitches, we use time coincidence.
Hence, enforcing time coincidence increases $P_{inj}$ of \texttt{Injections} (top panel), while completely discarding the glitches (bottom panel). Moreover, under the assumption that tracks with a low number of triggers are produced by detector noise we limit our analysis to tracks with 10 triggers or more.

Time coincidence, together with track reduction, discards many \texttt{Injection} tracks: while we have $21900$ tracks in H1 and $43100$ tracks in L1, their coincidence yields only $8967$ tracks. As before, in Appendix \ref{sec:testl1v1} we present the coincident results of the pairs L1 and V1, and H1 and V1, where an identical behavior can be observed.

\subsection{\texorpdfstring{Diving into the \textit{known data}: significance of $P_{inj}$ statistic}{Diving into the \textit{known data}: significance of Pinj statistic}}\label{sec:performance_gw}

\begin{figure}[]
    \centering
    \includegraphics[width=0.5\textwidth]{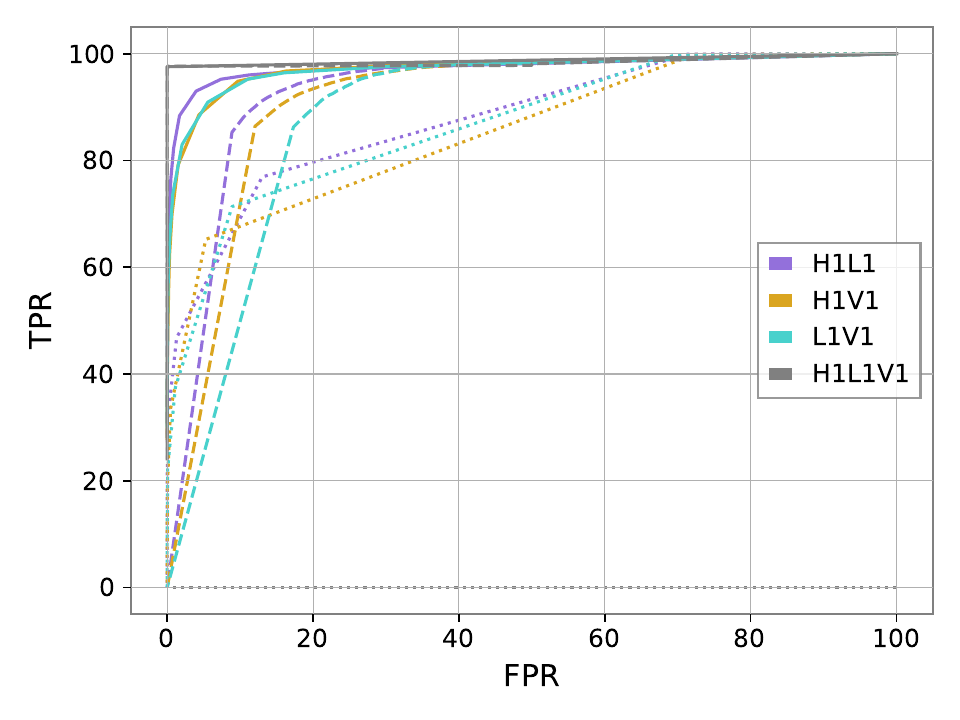}
    \caption{ROC curve of double and triple detector time (H1L1V1) for Eq. \ref{eq:old_stat} (dashed), Eq. \ref{eq:new_stat} (solid) and Eq. \ref{eq:mf_stat} (dotted). Here, TPR is the rate of correctly classified GW injections from the zero lag, while FPR is the rate of incorrectly classified time slides background tracks.}
    \label{fig:roc_time_slides}
\end{figure}

Up until now, we performed the standard statistical tests in the field of ML with $P_{inj}$ as our {classification} probability. Nonetheless, a ML based detection algorithm requires a proper understanding of the significance of its statistics considering the accidental background. With the time-shifted background, we intend to find extreme values of $\bar{P}_{inj}$ (see Eq.~\ref{eq:harmonic_mean}), but due to the probabilistic nature of ML algorithms, our statistic saturates at 1 as we can see in  Fig. \ref{fig:coincidence}. Since we are interested in the region with the highest probability we redefine our statistics as

\begin{equation}
\label{eq:old_stat}
- \log{(1-\bar{P}_{inj}) + \epsilon}, 
\end{equation}
where $\bar{\cdot}$ represents again the harmonic mean (see Eq.~\ref{eq:harmonic_mean} for a definition), and $\epsilon = 10^{-20}$ is added to prevent undefined behavior when $\bar{P}_{inj} \approx 1$, ensuring numerical stability, firstly proposed in \cite{Trovato:2023bby}. The main advantage of using the harmonic mean in this context is that it places more emphasis on smaller values, which helps to down-rank candidates with lower statistics. To assess the performance of this statistic we compute the ROC curve, but with a different definition of TP and FP. TP are coincident GW injections of O3a present in the zero lag that has been identified by GstLAL, i.e. they produce a track. On the other hand, FP are the coincident tracks of the time shifts of O3. With this definition we can construct the ROC curve stepping on different values of our statistic $- \log{(1-\bar{P}_{inj}) + \epsilon}$. Note that better performing ROC would be the ones that maximize the area under the curve (AUC). It is relevant to note that, as we are working with triple time if one detector does not observe the signal, then its contribution $P^{-1}_{inj, k} = 0$ (see Eq. \ref{eq:harmonic_mean}).

\begin{table}
\centering
\caption{Area under the curve (AUC) from ROC curves in Fig. \ref{fig:roc_time_slides}. An ideal classifier will have a $10^4$ AUC score.
}\label{tab:auc_old_new}
\begin{tabular}{@{}ccc@{}}
\toprule
\textbf{} & \diagbox[width=3cm, height=1.5cm]{Detector}{Time} & \textbf{H1L1V1} \\ \midrule
\multicolumn{1}{c|}{\multirow{4}{*}{\rotatebox[origin=c]{0}{\textit{$- \log{(1-\bar{P}_{inj})}$}}}} 
                                            & \textit{H1L1}   & 9281.54 \\
\multicolumn{1}{c|}{}                       & \textit{H1V1}   & 9113.69 \\
\multicolumn{1}{c|}{}                       & \textit{L1V1}   & 8843.81 \\
\multicolumn{1}{c|}{}                       & \textit{H1L1V1} & 9838.81       \\ \midrule \hline
\multicolumn{1}{c|}{\multirow{4}{*}{\rotatebox[origin=c]{0}{\textit{$\overline{P_{inj}/(\xi^2/\text{SNR}^2)}$}}}} 
                                            & \textit{H1L1}   & 9752.80 \\
\multicolumn{1}{c|}{}                       & \textit{H1V1}   & 9716.73 \\
\multicolumn{1}{c|}{}                       & \textit{L1V1}   & 9712.19 \\
\multicolumn{1}{c|}{}                       & \textit{H1L1V1} & 9877.87
\\ \midrule \hline
\multicolumn{1}{c|}{\multirow{4}{*}{\rotatebox[origin=c]{0}{\textit{$\overline{\rho_{\text{MF}}}$}}}}
                                            & \textit{H1L1}   & 8795.28 \\
\multicolumn{1}{c|}{}                       & \textit{H1V1}   & 8548.83 \\
\multicolumn{1}{c|}{}                       & \textit{L1V1}   & 8686.17 \\
\multicolumn{1}{c|}{}                       & \textit{H1L1V1} & 0.00  \\ \bottomrule
\end{tabular}
\end{table}

In Fig. \ref{fig:roc_time_slides} we present the ROC curve of $- \log{(1-\bar{P}_{inj}) + \epsilon}$ (dashed lines) in triple detector time (H1L1V1) for different detector combinations.  \mlp{We can observe that H1L1 has a better performance than H1V1 and L1V1. This could be due to a known lower sensitivity of V1 detector. However, L1V1 is worse performing than H1V1. A possible explanation for this is that L1 contains more glitches due to its higher sensitivity, confussing our ML model.} 

On the other hand, H1L1V1 has a better performance than double-coincidences since we only have two background tracks. We summarize their AUC in Table \ref{tab:auc_old_new}.

As we have seen for L1V1, it is possible that the behavior of MLP is overly optimistic, caused by a combination of its training in a completely controlled population-GW simulations and well-known glitches-and an overly simplified track input. Because of this, we explored a combination of $P_{inj}$ with variables used in traditional GW searches, such as SNR and $\xi^2$. In this work, the best-performing statistic {is found to be} was
\begin{equation}
\label{eq:new_stat}
\overline{P_{inj}/(\xi^2/\text{SNR}^2)},
\end{equation}
where again $\bar{\cdot}$ represents the harmonic mean. We present in Fig. \ref{fig:roc_time_slides} (solid lines) its ROC curve in triple detector time (H1L1V1). We can observe that all double coincidences have improved greatly with respect to the statistic $- \log{(1-\bar{P}_{inj}) + \epsilon}$. Moreover, in triple coincidence, the AUC is 9877.87, while before it was 9838.81 (see Table \ref{tab:auc_old_new}).

\mlp{To further understand the performance of our ML-enhanced statistic compared to standard matched filtering, we compare it to an approximate detection statistic defined in \cite{Magee:2023muf} and \cite{2024arXiv241015513M}} as

\begin{equation}
\label{eq:mf_stat}
\rho_{\text{MF}} = \frac{\rho}{[\frac{1}{2}(1 + \max{(1, \xi^{2})}^3)]^{1/5}}
\end{equation}
\mlp{for a single detector. As before, we use the harmonic mean to combine the different interferometers, yielding $\overline{\rho_{\text{MF}}}$ statistic. The resulting ROC curves are shown in  Fig. \ref{fig:roc_time_slides} (dotted). We can see that  $\rho_{\text{MF}}$ outperforms $- \log{(1-\bar{P}_{inj}) + \epsilon}$ at low FPR, but it has a worse performance than $\overline{P_{inj}/(\xi^2/\text{SNR}^2)}$. This can also be observed in Table \ref{tab:auc_old_new}, where all the values of $\overline{\rho_{\text{MF}}}$ are lower than $\overline{P_{inj}/(\xi^2/\text{SNR}^2)}$. It is interesting to note that $\overline{\rho_{\text{MF}}}$ for three detectors is zero, which might be caused due to the lack of sufficient background statistic for this combination of detectors.}

\begin{figure}[]
    \centering
    \includegraphics[width=0.5\textwidth]{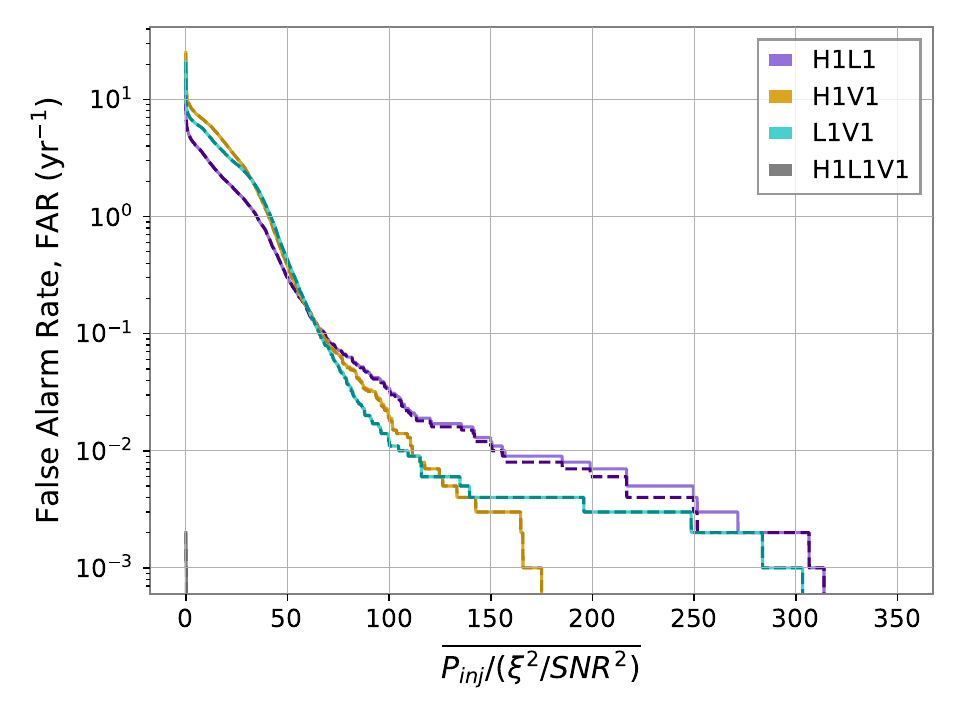}
    \caption{False alarm rate (FAR) as a function of the ranking statistic in triple time (H1L1V1) for different detector combinations. The solid lines show the time-shifted background with real GW signal tracks, while dashed lines show the time-shifted background without them.}
    \label{fig:ifar_new_stat}
\end{figure}

False alarm rate (FAR) is typically used to assess the performance of a GW search. This is measured by time-shifting the detectors' data relative to each other, as described in Sec.~\ref{sec:timeshift}. Figure~\ref{fig:ifar_new_stat} shows the FAR as a function of our ranking statistic $\overline{P_{inj}/(\xi^2/\text{SNR}^2)}$ for different detector combinations in H1L1V1 time. Note that H1L1V1 is in the bottom left corner as it has two samples.

A principal characteristic of a GW ranking statistic is that it increases monotonically as the probability of finding a GW signal increases. Because of this, real GW signal tracks would lay on the far right of the FAR distribution. To investigate whether our ranking statistic behaves in this way we remove all tracks that correspond to real GW signals, yielding the dashed line distributions. While there is a noticeable discrepancy in the H1L1 distribution, this issue does not occur with H1V1 and L1V1. This may be due to the lower sensitivity of V1 or a limitation of our statistics.

\begin{figure}[!h]
\subfloat[\label{fig:tpr_snr}{ TPR as a function of injected SNR}.]{%
\includegraphics[width=0.45\textwidth]{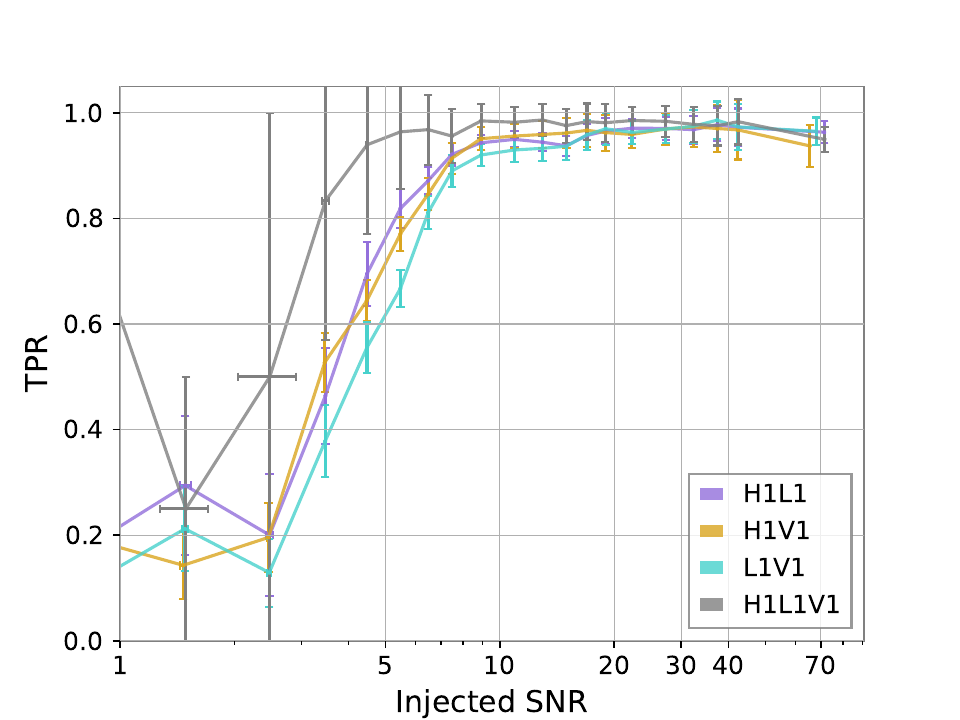}%
}\hfill
\subfloat[\label{fig:tpr_dist}{TPR as a function of luminosity distance}. ]{%
\includegraphics[width=0.45\textwidth]{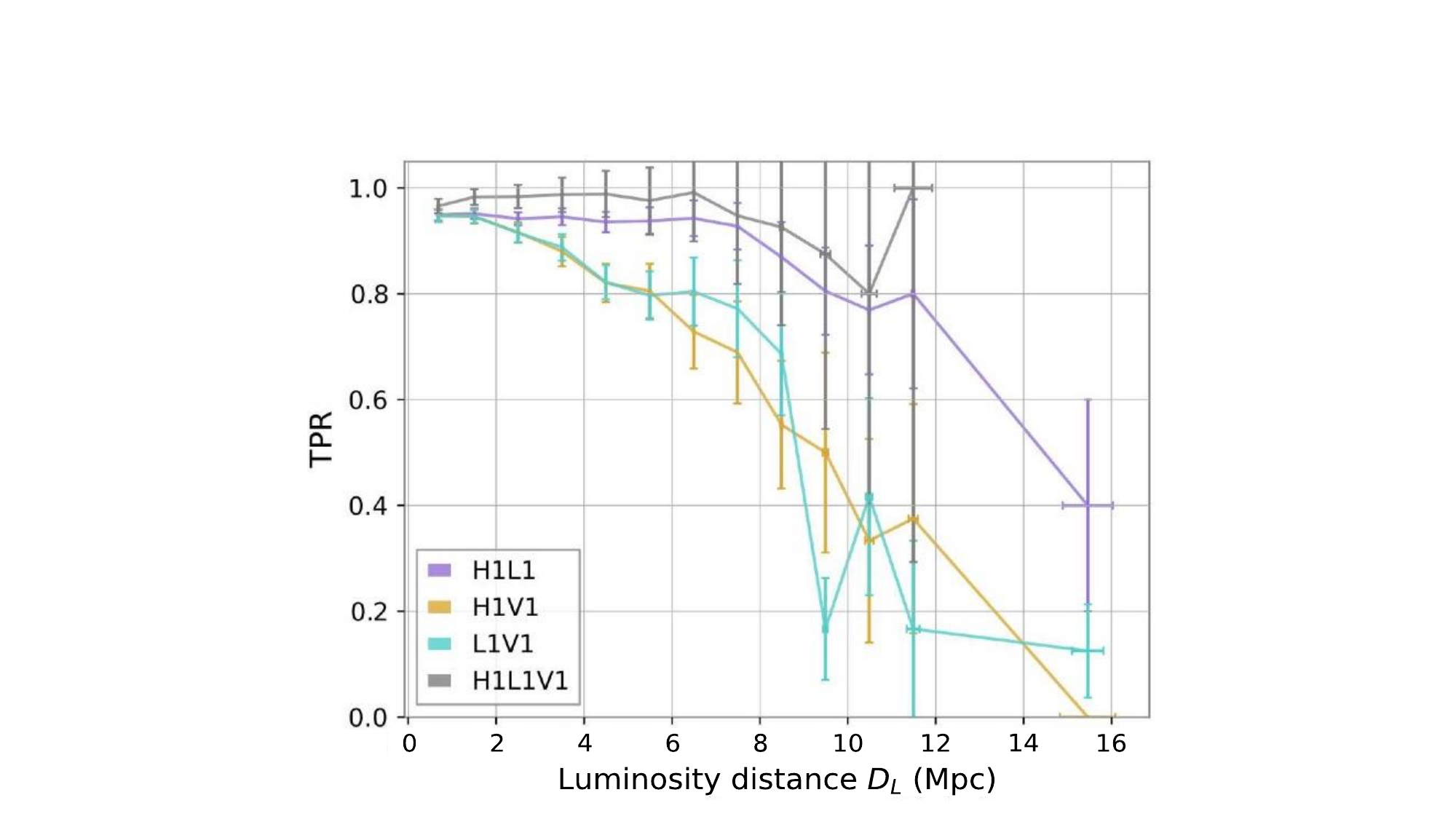}%
}\hfill
\caption{TPR, i.e., the rate of correctly classified GW injections, at FAR = 0.1 (equivalent to one per 10 years) for different detector combinations in triple time (H1L1V1). Error bars represent the standard error at 3 standard deviations.}
\label{fig:tpr_new_stat}
\end{figure}

To further understand the performance of our statistic, we measure the recovery of GW injections from O3a. To compute the TPR, we fix FAR at 0.1, or equivalently, one per 10 years, using the results in Fig. \ref{fig:ifar_new_stat}. In Fig. \ref{fig:tpr_new_stat} we present TPR as a function of the injected SNR (Fig. \ref{fig:tpr_snr}) and as a function of the luminosity distance $D_{L}$ in Mpc (Fig. \ref{fig:tpr_dist}).

In Fig. \ref{fig:tpr_snr} we can observe that for all detector combinations TPR $\approx 0.95$ for injected SNR $\gtrsim 8$. For SNR $\lesssim 8$ there is a drop in performance for double coincidence tracks, while triple coincident tracks  maintain their TPR until $\approx 5$ injected SNR. However, we must note that for an injected SNR $\approx 5$ there is less statistic, as we can observe from the increase in size of the vertical error bars.  In Fig. \ref{fig:tpr_dist} we can observe that the TPR for H1V1 and L1V1 decreases steeply for $D_{l} \gtrsim 0.6\,$Mpc, possibly due to the lower sensitivity of V1. Moreover, {H1L1} and {H1L1V1} have a similar trend, with a drop in performance at $D_{l} \gtrsim 1.2\,$Mpc. As in the case of the injected SNR, there is less statistic for $D_{l} \gtrsim 1\,$Mpc. In future work, we could conduct a broader injection campaign to mitigate these issues.

\section{Discussion}

In the previous Sections, we showcased the utilization of matched-filtering triggers to learn the patterns, or tracks, of different types of signals from single-GW detectors. It is relevant to highlight the reliance of the ML model on a mere 6 parameters (see Eq. \ref{eq:feature}) to perform the multi-classification task. This approach does not only recover successfully the test set of the population we have trained on but also accurately recovers injections of GW signals. In future works, it would be interesting to compare these results with state-of-the-art algorithms to better quantify the performance of our method. 

Despite these achievements, certain limitations must be noted. The primary constraint is that we characterize the feature vector of our input with only 6 features. Before this investigation, we have not utilized a feature selection procedure for the definition of the input feature vector, as we have limited ourselves to variables that are well-understood in the GW community. Such an approach could enhance the performance of the model. 

Another limitation pertains to the ranking statistic itself. In this work, we have combined the statistic of the ML model with standard measurements such as SNR and $\xi^2$. Due to time constraints, this exploration was limited, but it does not imply that our ranking statistic is the optimal one. In future works we will further explore enhancing our statistic with different variable combinations. However, we expect that the largest improvement would come from the inclusion of the time dimension, i.e. the behavior of the signal through the template bank. This will most likely provide valuable insight useful to enhance the distinction between these classes. 

\section{Conclusion}
\label{sec:conclusions}

In this investigation, we propose a flexible method to detect CBC signals combining the robustness of matched-filtering as an optimal filter, with the generalization power of ML algorithms.  Thus, we construct an MLP model to learn from sets of the matched filtering triggers, labelled here as tracks, and perform a multi-classification task in a single detector. Specifically, we tackle the  IMBH search of GstLAL during O3, but this method could be extended to other CBC signals and matched-filtering algorithms.

Since multiple templates could potentially match a given signal, meaning that the  input to the model would have variable length, in this proof-of-concept work, we reduced the dimensionality of the problem by averaging the matching templates and weighting them by SNR. Another difficulty of our task is to deal with highly unbalanced data, so to mitigate this problem, we undersample large classes and oversample small classes. Nonetheless, this could bias the model towards certain repetitive features, so to avoid overfitting we employ 9-fold-cross-validation and early-stopping algorithm. In the following, we revisit some of the major results of this method:

\begin{itemize}
\item[-] \textit{Controlled data set:} We trained our model in O3a data, and we tested its generalization ability in O3b. We used the standard  statistics to test its robustness. We obtained accuracies of $98.8\%, 95.8 \%$ and $99.3 \%$ for H1, L1 and V1 in O3a test set, while showing a sharp decrease in accuracy of $73.0\%, 67.5 \%$ and $75.0 \%$ for H1, L1 and V1 in O3b test set (see Section \ref{sec:performance}). This drop in performance was caused by the influence of glitches, so we decided to implement time coincidence to lower this background.

\item[-] \textit{Novel time coincidence:} To enhance the performance of the model, we enforce time coincidence among detectors, which greatly reduces the background of glitches. As we are dealing with clusters of triggers, instead of single triggers, we take the average time of the cluster and the light time travel between detectors to consider that tracks from different detectors coincide in time (see Section \ref{sec:time_coincident}).
\item[-] \textit{Computational efficiency:} This method only uses 6 variables to perform this classification (see Eq. \ref{eq:feature}), and while its training is intensive, we can classify a given input in $2.9 \times 10^{-6}\,$s. As this process is highly parallelizable, this computation essentially is inexpensive.
\item[-] \textit{Construction of accidental background}: We constructed the time shift background to assess the significance of our ranking statistic (see Section \ref{sec:performance_gw}) and tested its performance on simulated IMBH signals. With an FAR $= 0.1\,$yr, we have a TPR $\gtrsim 0.8$ at SNR $\sim 5$ for all detector combination. Regarding the luminosity distance, H1L1 
and H1L1V1 have a TPR $\gtrsim 0.8$, with a drop in performance at $D_{L} \gtrsim 1.2\,$Mpc. 

\end{itemize}

In summary, this method has shown not only to have a robust performance in a controlled environment classifying IMBH injections but also shown its generalization power in finding simulated GW signals, demonstrating that it is possible to form a synergistic relationship between current state-of-the-art matched-filtering techniques and novel ML methods.

In future work, we will explore a different ML method that can process varying length inputs, as the time information might be relevant to enhance the performance of the model. For this aim, several ML algorithms could be employed, such as recurrent neural network \cite{lstm} or even transformers with an attention mechanism \cite{attention}. With such a model we could perform a fair comparison with state-of-the-art pipelines to quantify the potential improvements in IMBH searches that this technology may offer. Furthermore, and as we mentioned before, this  methodology is flexible and simply relies on matched filtering computation. We could therefore extend it to other CBC signals and state-of-the-art matched-filtering algorithms.

\section{Acknowledgements}\label{sec:acknowledgements}
We would like the thank Harsh Narola for his useful comments. S.C. is supported by the National Science Foundation under Grant No. PHY-2309332. C.C. acknowledges support from NSF Grant No.~PHY-2309356. This project was supported by Nikhef Laboratory, and we would like to thank the Nikhef Computing group. The authors are also grateful for computational resources provided by the LIGO Laboratory and supported by the National Science Foundation Grants No. PHY-0757058 and No. PHY-0823459. This material is based upon work supported by NSF’s LIGO Laboratory which is a major facility fully funded by the National Science Foundation.
\label{s:acknow}

\bibliography{references}

\onecolumngrid
\appendix
\section{Results of LIGO Livingston and Virgo}
\label{sec:appendix}

\subsection{Selecting a time window}\label{sec:twl1v1}

\begin{figure}[h]
\subfloat[\label{fig:twl1o3a}{ Testing on O3a with \textit{known data set}}.]{%
\includegraphics[width=0.5\columnwidth]{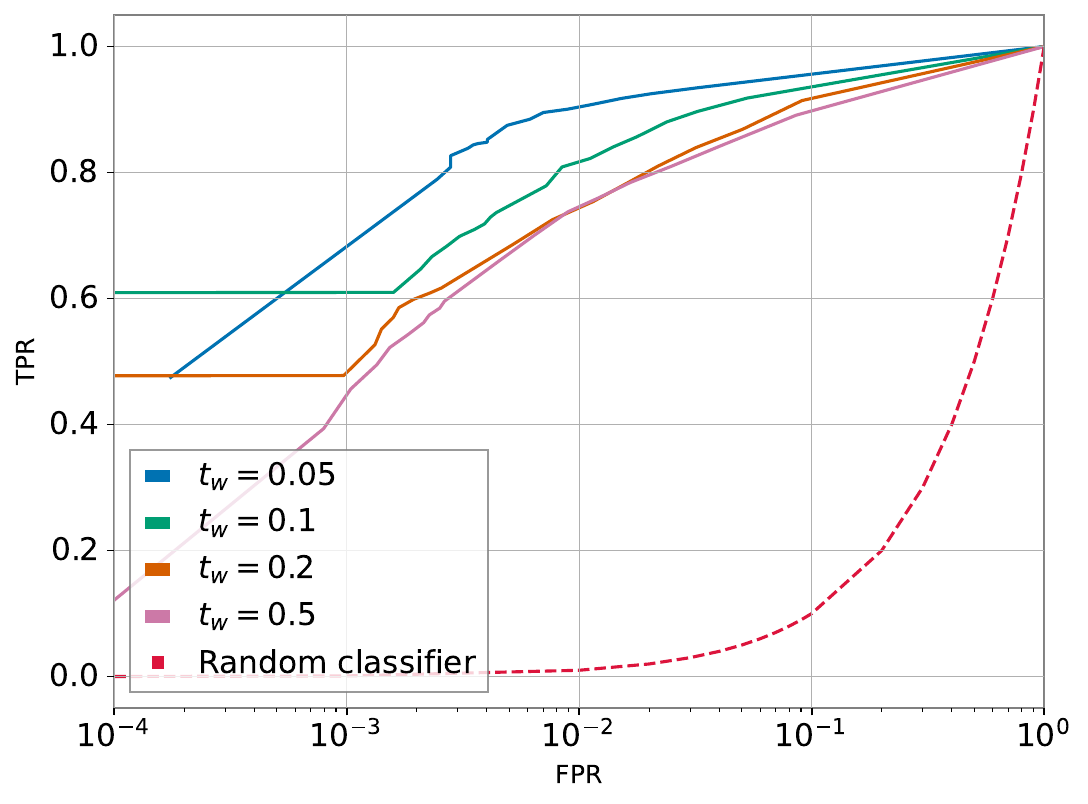}%
}\hfill
\subfloat[\label{fig:twl1o3b}{Testing on O3b with \textit{known data set}}.]{%
\includegraphics[width=0.5\columnwidth]{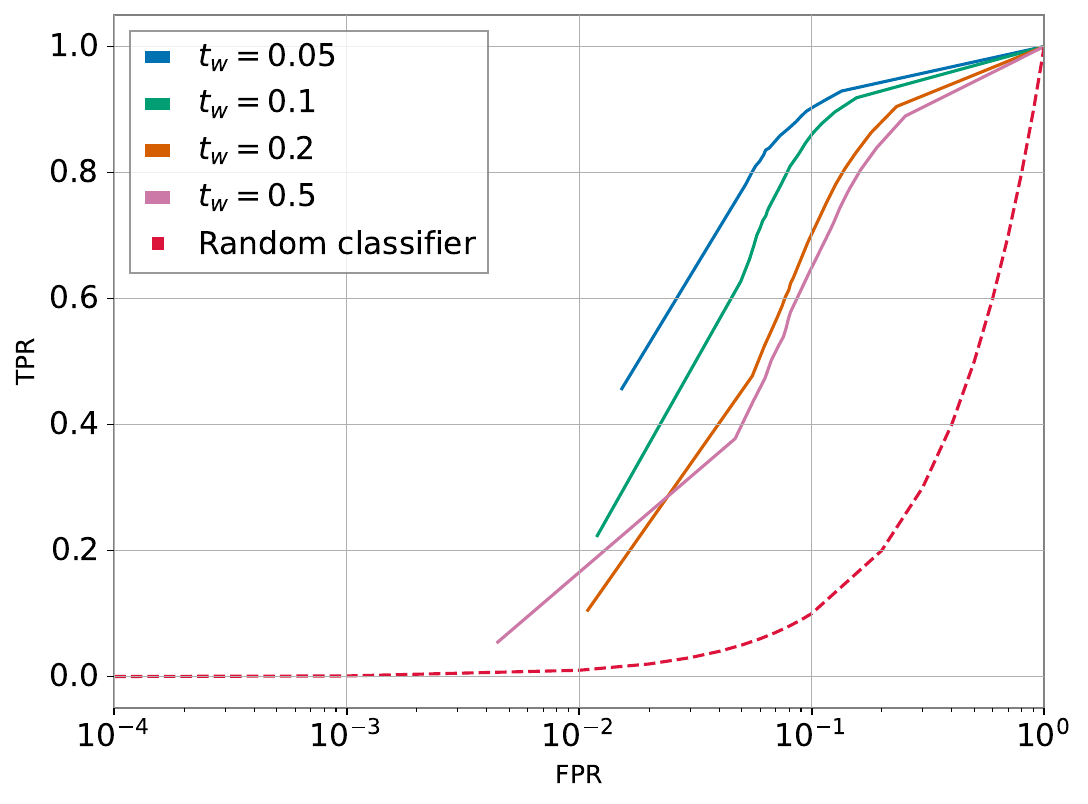}%
}\hfill
\caption{ROC curve for different time windows $t_{w}$ in L1, i.e. TPR  as a function of False Positive Rate FPR in logarithmic scale. \textit{(Left)} Testing in the \textit{known data set} of O3a. \textit{(Right)} Testing in the \textit{known data set} of O3b. Note that the dashed line indicates a random guess.}
\label{fig:tunetwL1}
\end{figure}

\begin{figure}[h]
\subfloat[\label{fig:twv1o3a}{ Testing on O3a with \textit{known data set}}.]{%
\includegraphics[width=0.5\columnwidth]{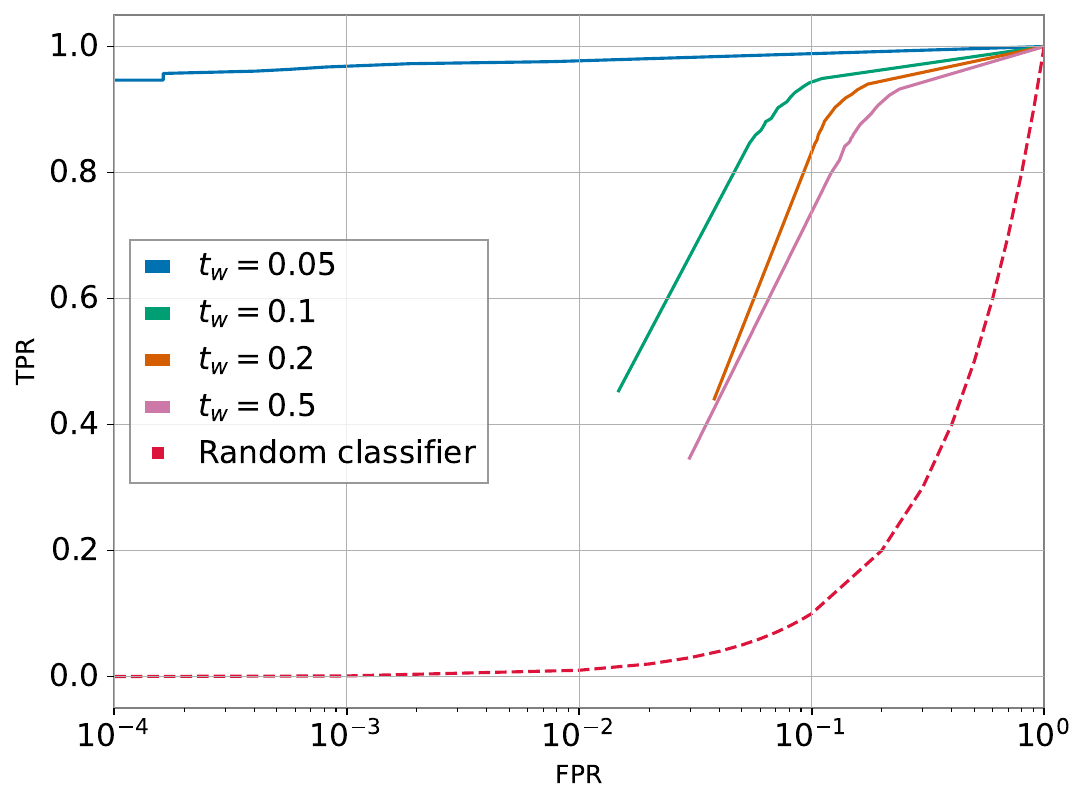}%
}\hfill
\subfloat[\label{fig:twv1o3b}{Testing on O3b with \textit{known data set}}.]{%
\includegraphics[width=0.5\columnwidth]{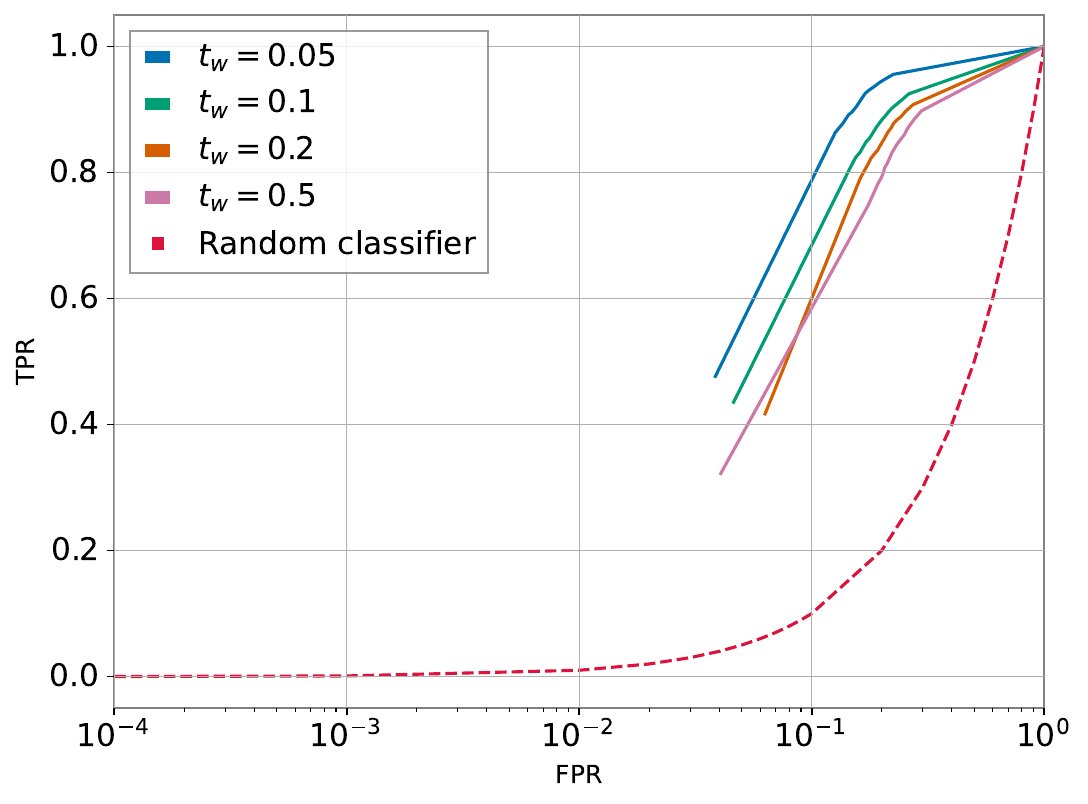}%
}\hfill
\caption{ROC curve for different time windows $t_{w}$ in V1, i.e. TPR  as a function of False Positive Rate FPR in logarithmic scale. \textit{(Left)} Testing in the \textit{known data set} of O3a. \textit{(Right)} Testing in the \textit{known data set} of O3b. Note that the dashed line indicates a random guess.}
\label{fig:tunetwV1}
\end{figure}

Similarly to Section \ref{sec:tw}, we show in Fig. \ref{fig:tunetwL1} and Fig. \ref{fig:tunetwV1} the ROC curves of L1 and V1, respectively.  We can observe that the TPR degrades as we increase the size of $t_{w}$, being this decrease sharper for the \textit{known data set} of O3b than for the  \textit{known data set} of O3a.  Notably, for V1 (see Fig. \ref{fig:tunetwV1}), while the ROC curve of $t_{w}=0.05$ in O3a is almost constant, the ROC curve of O3b decreases even faster than in the case of O3b L1.

As in Fig. \ref{fig:tunetwH1}, in Fig. \ref{fig:tunetwL1} and Fig. \ref{fig:tunetwV1} $t_{w} = 0.05\,$s has a better performance than other time windows. A possible explanation for this behavior could be that since IMBH signals are short, larger time windows would add random triggers that are unrelated to the IMBH signal itself, biasing the model. Hence, as in Section \ref{sec:tw} we conclude that $t_{w}=0.05$ is the best time window for our task.

\subsection{Training with \textit{known data set} of O3a}
\label{sec:trainl1v1}

\begin{figure}[h]
\subfloat[\label{fig:lossesl1}{Ligo Livingston}.]{%
\includegraphics[width=0.5\columnwidth]{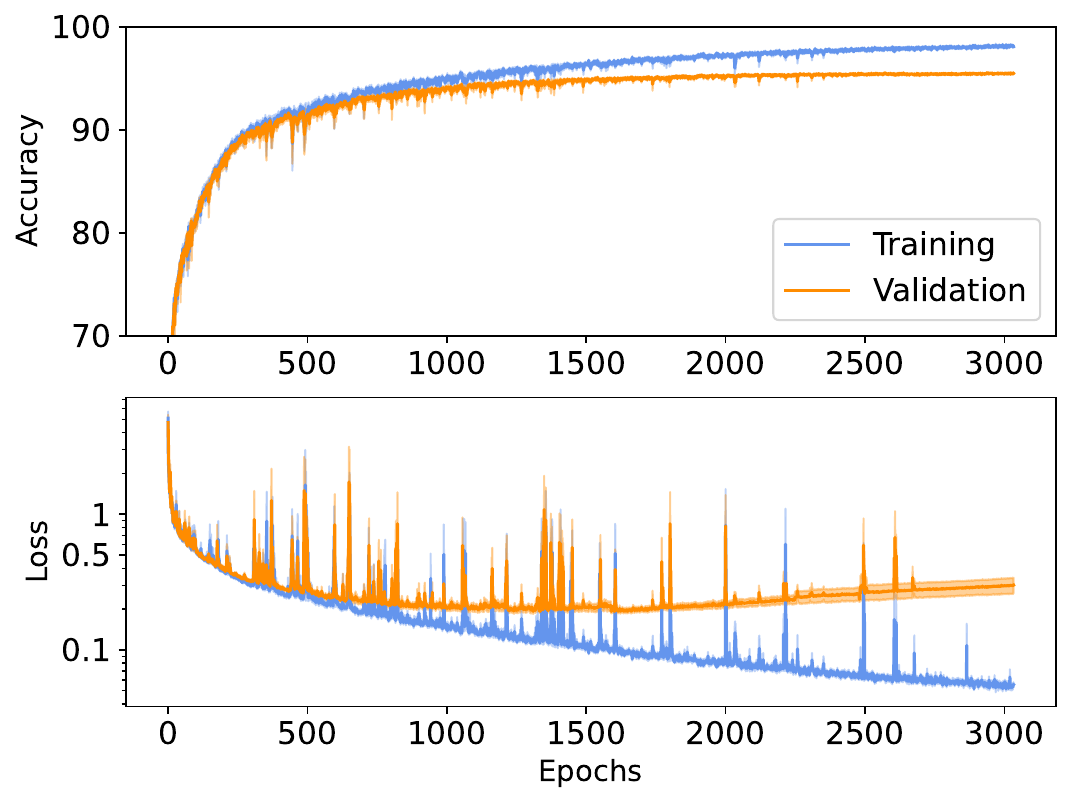}%
}\hfill
\subfloat[\label{fig:lossesv1}{Virgo}.]{%
\includegraphics[width=0.5\columnwidth]{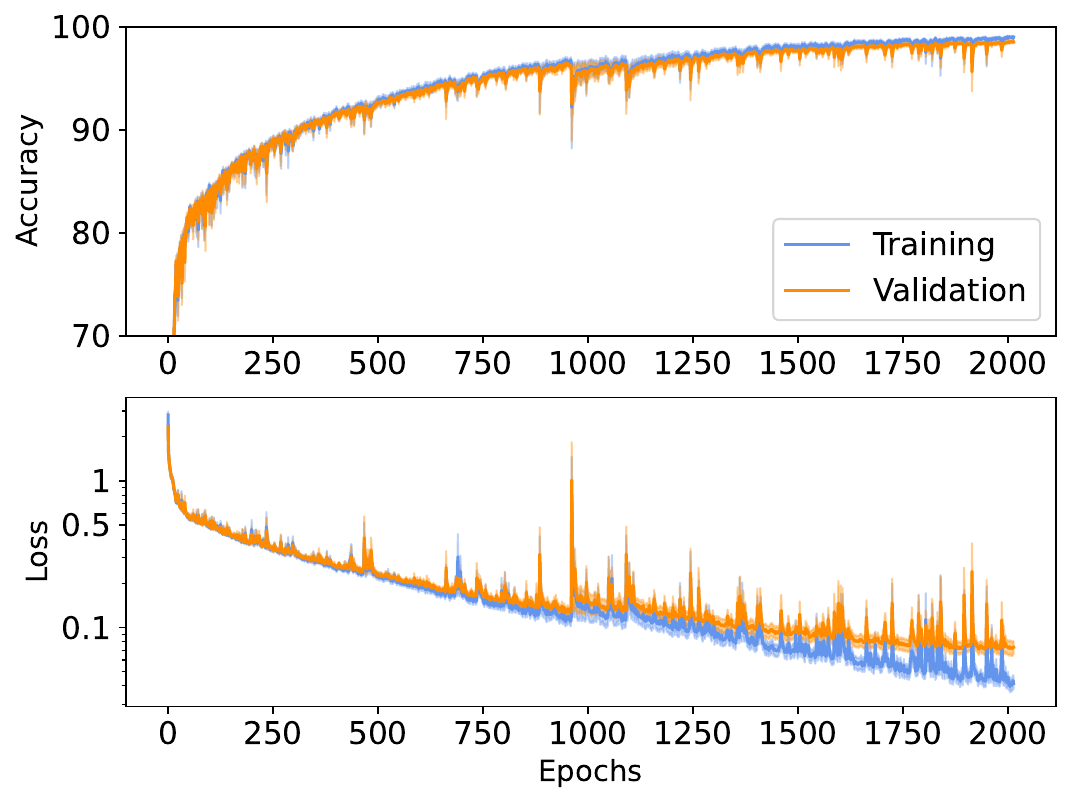}%
}\hfill
\caption{Comparison between training and validating with 9-fold cross-validation during training and testing. \textit{(Top row)} Mean accuracy at 3 standard deviations as a function of the epochs during. \textit{(Bottom row)} Average 9-fold cross-validation loss as a function of the epochs.}
\label{fig:cfmV1}
\end{figure}

As in Section \ref{sec:train}, in Fig. \ref{fig:cfmv1} we show, for training and validation, the mean accuracy (top row) and loss (bottom row) of the 9-fold cross-validation as a function of the epochs for L1 (left column) and V1 (right column), where the shadowed region represents ±3 standard deviation. Note that the loss is plotted in logarithmic scale, so we can appreciate that there is less overfitting in V1 than in L1.

Another relevant point is the peaks in the loss functions, in particular around epoch 500 in L1 and around epoch 1000 in V1. These peaks indicate a change in the learning rate due to the adaptive learning rate scheme (see Section \ref{sec:mlp} for details).

\subsection{Diving in to the  \textit{known data}: performance evalutation}
\label{sec:testl1v1}

\begin{figure}[h]
\subfloat[\label{fig:cfml1o3a}{ O3a \textit{known data set}}.]{%
\includegraphics[width=0.5\columnwidth]{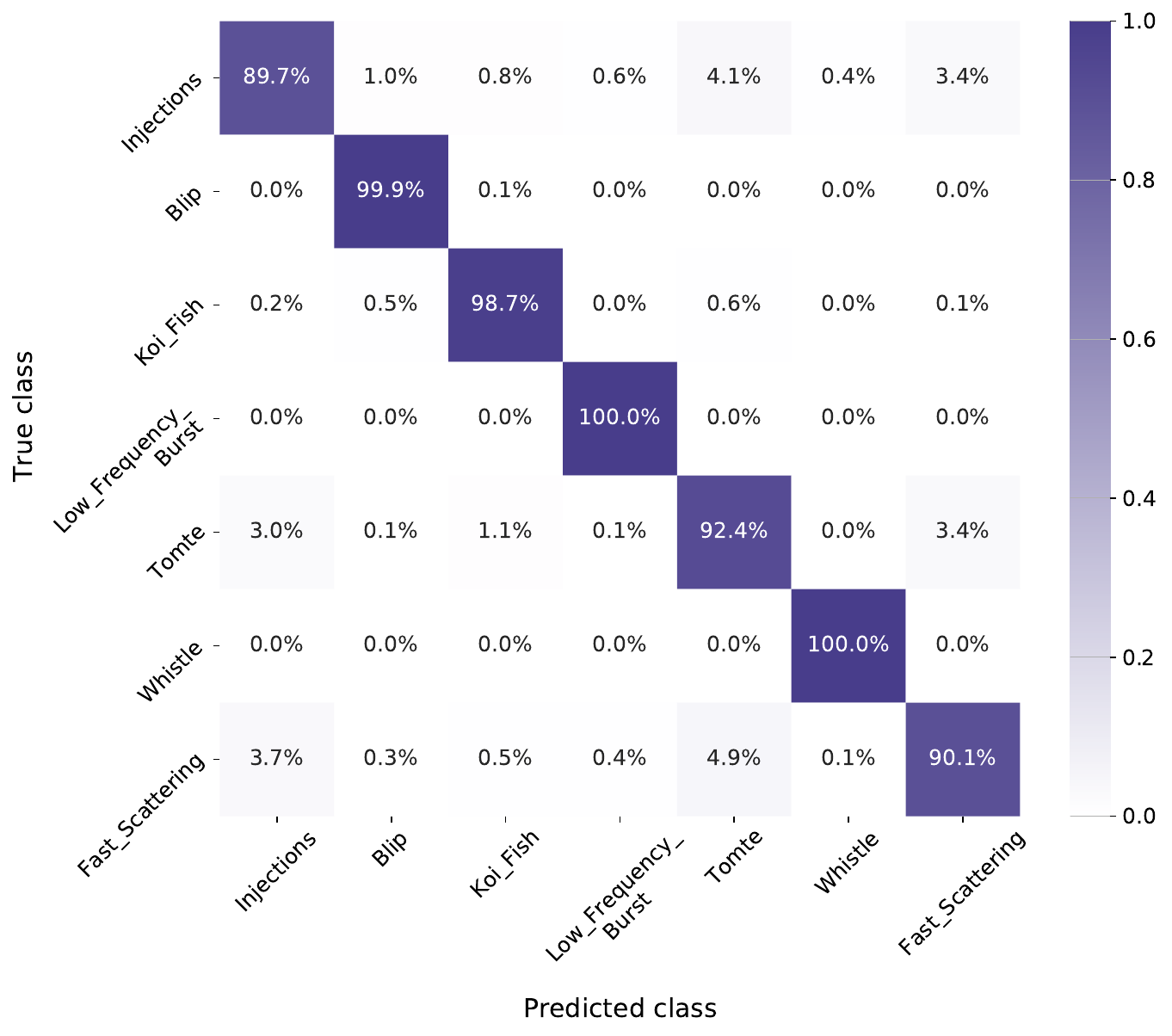}%
}\hfill
\subfloat[\label{fig:cfml1o3b}{O3b \textit{known data set}}.]{%
\includegraphics[width=0.5\columnwidth]{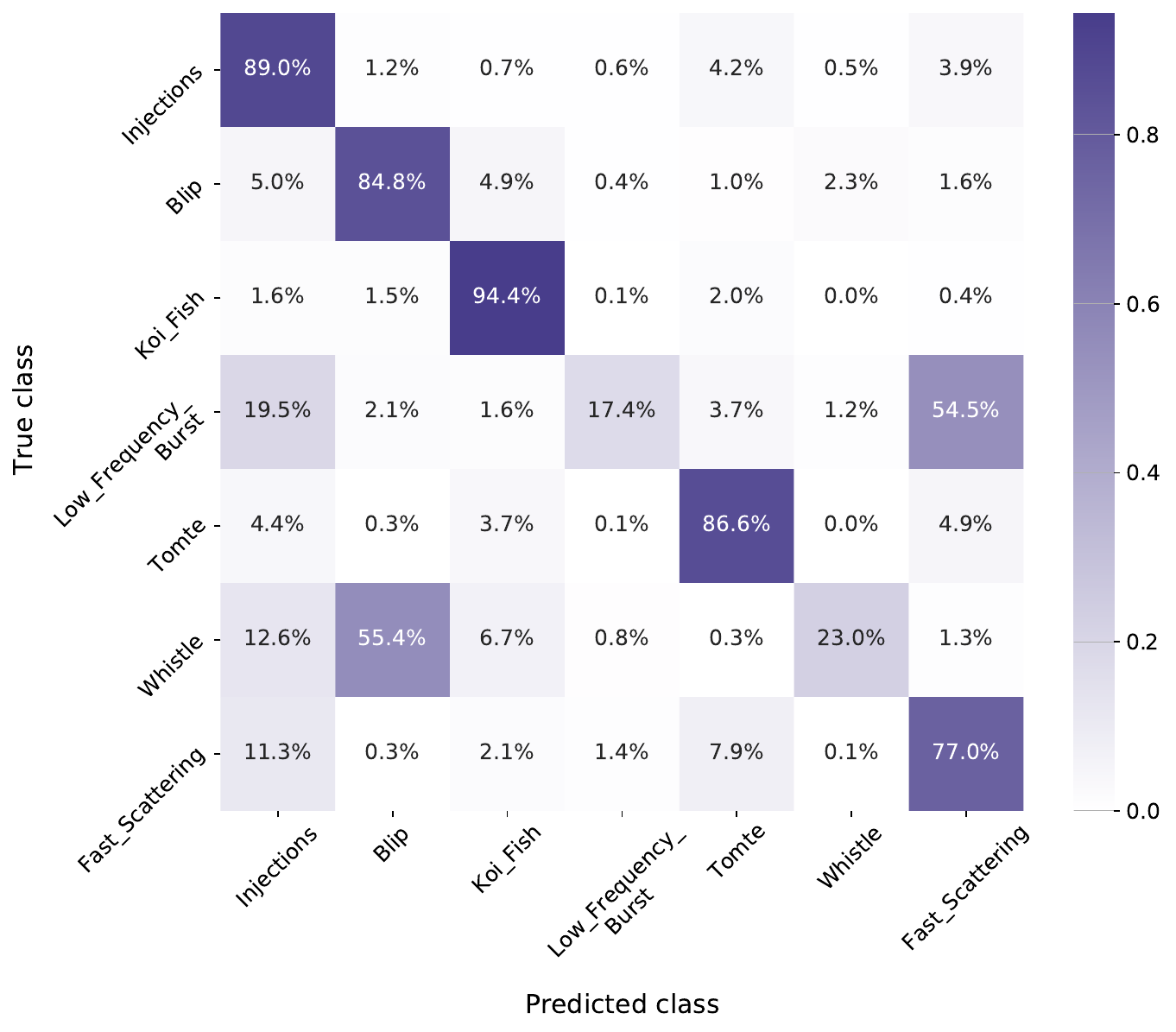}%
}\hfill
\caption{Relative values of the confusion matrix for the test set for L1.}
\label{fig:cfml1}
\end{figure}

\begin{figure}[!h]
\subfloat[\label{fig:cfmv1o3a}{ O3a\textit{known data set}}.]{%
\includegraphics[width=0.5\columnwidth]{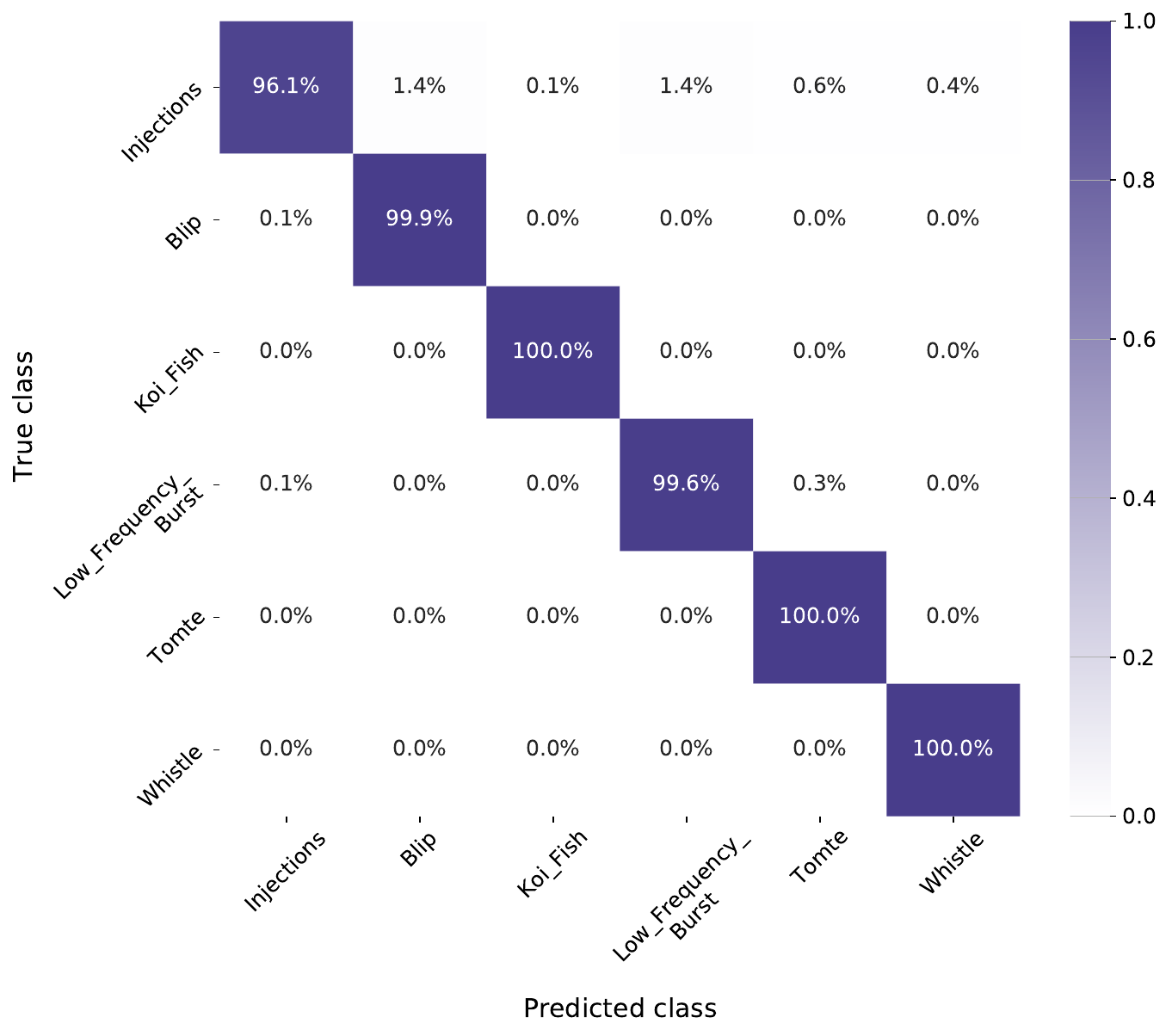}%
}\hfill
\subfloat[\label{fig:cfmv1o3b}{O3b\textit{known data set}}.]{%
\includegraphics[width=0.5\columnwidth]{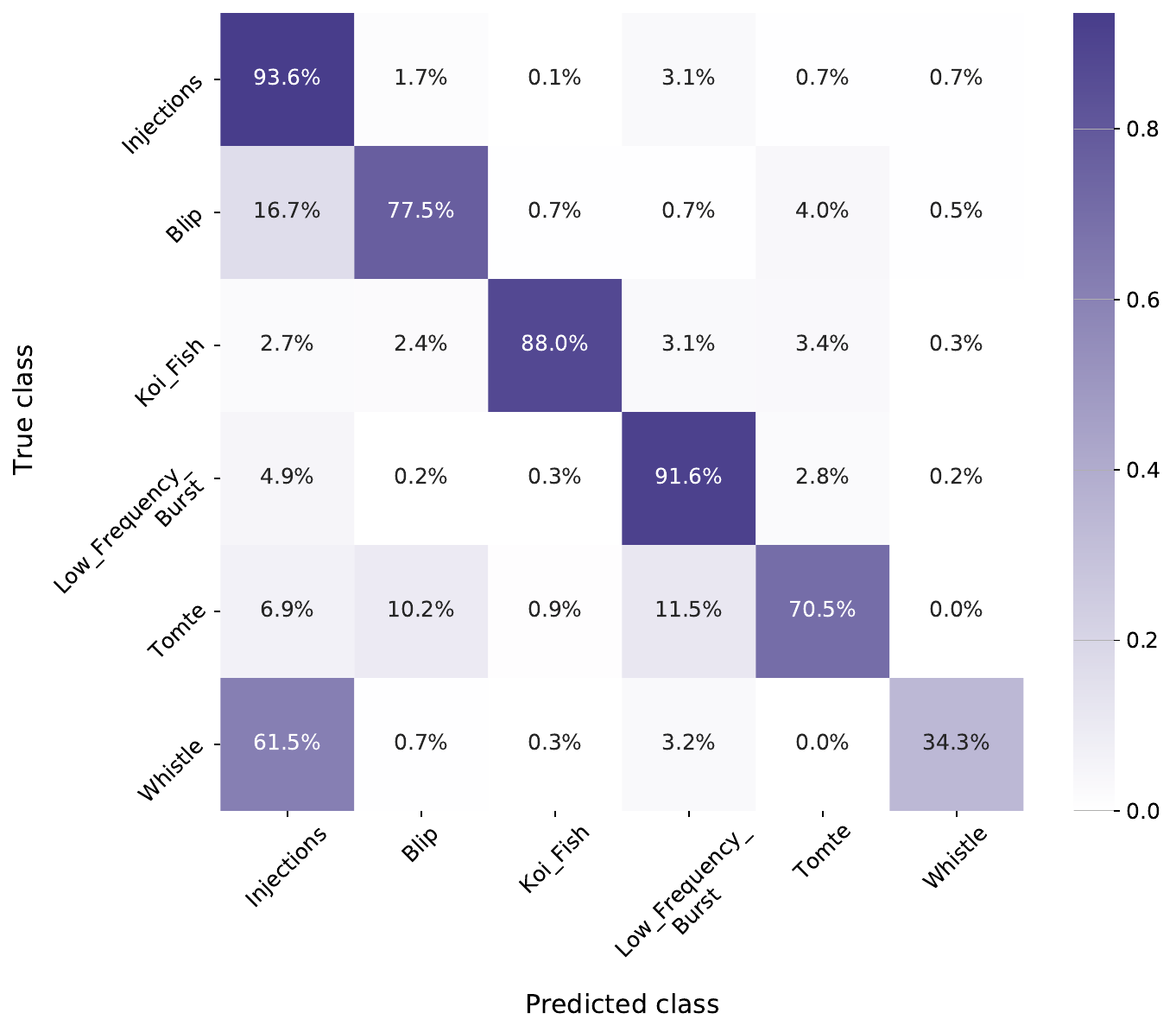}%
}\hfill
\caption{Relative values of the confusion matrix for the test set for V1.}
\label{fig:cfmv1}
\end{figure}

Similarly to Section \ref{sec:performance}, where we want to test the generalization power between the \textit{known data sets} of O3a and O3b for H1, we test  the \textit{known data sets} of O3a and O3b for L1 and V1, presenting their confusion matrices in Fig. \ref{fig:cfml1} and Fig. \ref{fig:cfmv1}, respectively. In Fig. \ref{fig:cfml1o3a} and Fig. \ref{fig:cfmv1o3a} we can see that most inputs are correctly classified, yielding an accuracy of $95.81\%$ in L1 and $99.27\%$ in V1. Nonetheless, Fig. \ref{fig:cfml1o3b} and Fig. \ref{fig:cfmv1o3b} we can observe an increase of false positives, decreasing the accuracy to $67.45\%$ in L1 and $75.91\%$ in V1.

In L1 (see Fig. \ref{fig:cfml1o3b}), the \texttt{Injections} are mostly correctly classified, with $1.2\%$ mislabelled as \textit{Blip}. However, $19.5\%$ of \texttt{Low\_Frequency\_Burst}, $12.6\%$ of \texttt{Whistle} and $11.3\%$ of \texttt{Fast\_Scattering} are incorrectly classified as \texttt{Injections}. Interestingly, $54.5\%$ of \textit{Low\_Frequency\_Burst} are misclassified as \texttt{Fast\_Scattering} since they share a similar frequency range.
In V1 (see Fig. \ref{fig:cfmv1o3b}), the  \texttt{Injections} are also mostly correctly classified, with $1.7\%$ mislabelled as \texttt{Blip}. However, $61.5\%$ of \texttt{Whistle} and $16.3\%$ of \texttt{Blip} are incorrectly classified as \texttt{Injections}. It is also interesting to note that in this detector only $4.9\%$ \texttt{Low\_Frequency\_Burst} are misclassified as \texttt{Injections}, while a $10.2\%$ and $11.5 \%$ of \texttt{Tomte} are incorrectly labelled as \texttt{Blip} or \texttt{Low\_Frequency\_Burst}, respectively.

\begin{figure}[!h]
\subfloat[\label{fig:coincidencel1v1}{ L1 (red) and V1 (green), and their coincidence (yellow)}.]{%
\includegraphics[width=0.5\columnwidth]{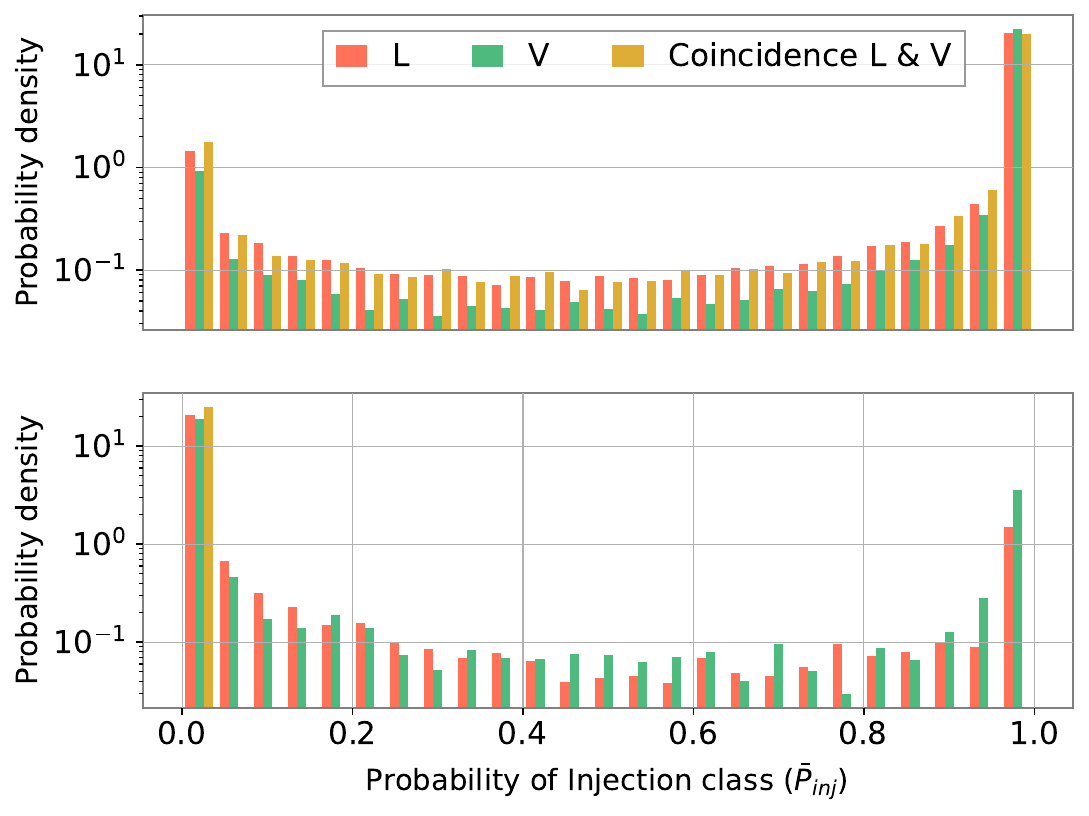}%
}\hfill
\subfloat[\label{fig:coincidenceh1v1}{H1 (blue) and V1 (green), and their coincidence (turquoise)}.]{%
\includegraphics[width=0.5\columnwidth]{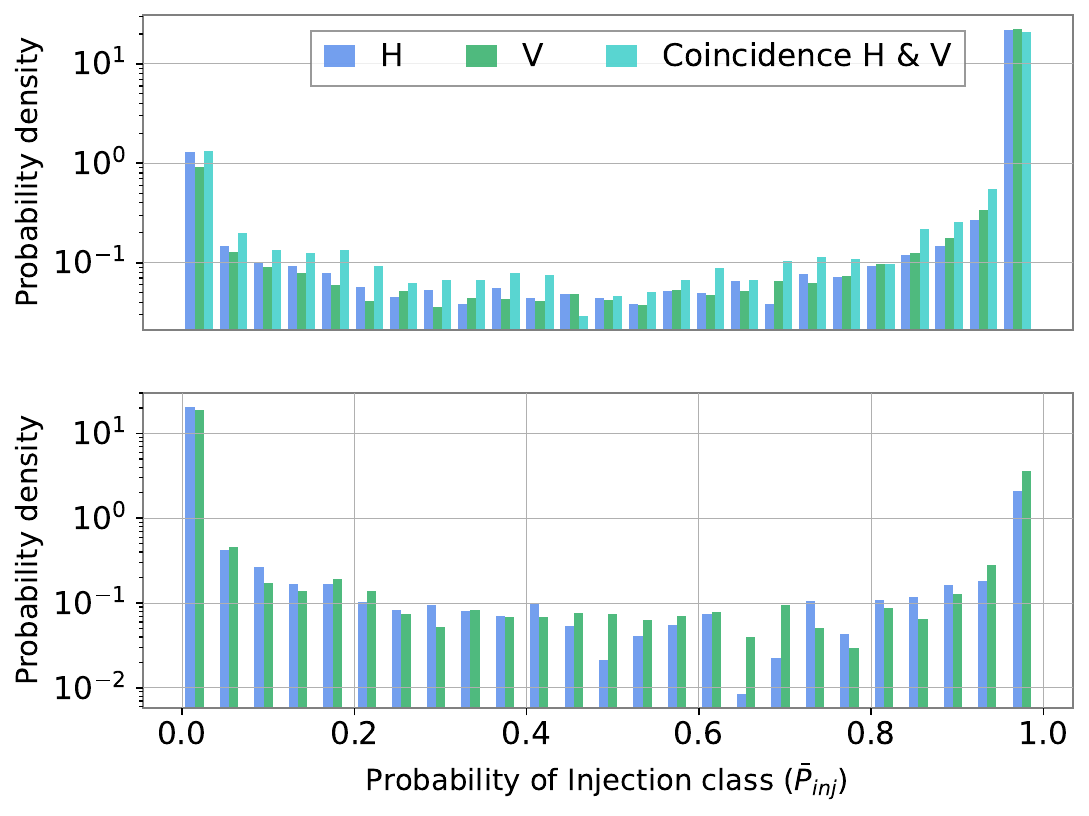}%
}\hfill
\caption{Probability density of being \texttt{Injection} ($P_{inj}$), using the \textit{known} test set of O3b.   \textit{(Top row)} Probability density of elements in the \texttt{Injection} class in logarithmic scale. \textit{(Bottom row)} Probability density of elements in all the other classes, i.e. glitches, in logarithmic scale. Given the counts of the $i$th bin $c_i$ and its width $b_i$, we define the probability density as $c_i/(\sum^N_i c_i \times b_i)$, where $N$ is the total number of bins of the histogram.}
\label{fig:coincidences}
\end{figure}

Due to the poor generalization between O3a and O3b, we need to reduce the number of false positives. For this aim, we enforce time coincidence as in Section \ref{sec:performance}. Similarly to Fig. \ref{fig:coincidence}, in Fig. \ref{fig:coincidences} we present the probability of being \texttt{Injection} ($P_{inj}$) for L1 (left column) and V1 (right column). In the top row, we show the probability density of $P_{inj}$ for the \texttt{Injection} class, and in the bottom row, we show the probability density of $P_{inj}$ for any other class. When we enforce time coincidence, represented in yellow for L1 and V1, and in turquoise for H1 and V1, $P_{inj}$ increases for the \texttt{Injection} class, while we completely discard all the other classes. Note that this is at the cost of reducing the number of  correctly classified \texttt{Injections}.

\end{document}